\documentclass[12pt,apj]{emulateapj}

\shortauthors{Reiners \& Basri}

\begin{document}


\title{A Volume-limited Sample of 63 M7--M9.5 Dwarfs\\
  II. Activity, magnetism, and the fade of the
  rotation-dominated dynamo}



\author{A. Reiners\altaffilmark{*}}
\affil{Institut f\"ur Astrophysik, Georg-August-Universit\"at,
  Friedrich-Hund-Platz 1, 37077 G\"ottingen, Germany}
\email{Ansgar.Reiners@phys.uni-goettingen.de}
\and
\author{G. Basri}
\affil{Astronomy Department, University of California, Berkeley, CA
  94720 }
\email{basri@berkeley.edu}


\altaffiltext{*}{Emmy Noether Fellow}


\begin{abstract}
  In a volume-limited sample of 63 ultracool dwarfs of spectral type
  M7--M9.5, we have obtained high-resolution spectroscopy with UVES at
  the Very Large Telescope and HIRES at Keck Observatory. In this
  second paper, we present projected rotation velocities, average
  magnetic field strengths, and chromospheric emission from the
  H$\alpha$ line. We confirm earlier results that the mean level of
  normalized H$\alpha$ luminosity decreases with lower temperature,
  and we find that the scatter among H$\alpha$ luminosities is larger
  at lower temperature. We measure average magnetic fields between 0
  and 4\,kG with no indication for a dependence on temperature between
  M7 and M9.5.  For a given temperature, H$\alpha$ luminosity is
  related to magnetic field strength, consistent with results in
  earlier stars. A few very slowly rotating stars show very weak
  magnetic fields and H$\alpha$ emission, all stars rotating faster
  than our detection limit show magnetic fields of at least a few
  hundred Gauss. In contrast to earlier-type stars, we observe
  magnetic fields weaker than 1\,kG in stars rotating faster than
  $\sim3$\,km\,s$^{-1}$, but we find no correlation between rotation
  and magnetic flux generation among them.  We interpret this as a
  fundamental change in the dynamo mechanism; in ultracool dwarfs,
  magnetic field generation is predominantly achieved by a turbulent
  dynamo, while other mechanisms can operate more efficiently at
  earlier spectral type.
\end{abstract}





\section{Introduction}

Ultracool dwarfs are objects of spectral types M7--M9.5. They can be
stars in the mass range 0.08--0.09\,M$_\Sun$ on the Main Sequence,
young brown dwarfs, or even very young objects of planetary mass. A
large fraction of ultracool dwarfs show strong chromospheric emission
\citep[e.g.,][]{West04} probably due to magnetic fields that are
sustained by a stellar dynamo \citep{RB07}.  At masses lower than
0.3\,M$_{\Sun}$ \citep{Burrows97}, ultracool dwarfs are fully
convective \citep{Baraffe98} and cannot harbor a dynamo that requires
a tachocline, as is believed to be the case at least for the cyclic
part of the solar dynamo \citep[e.g.,][]{Ossendrijver03}.

The relation between rotation and activity in solar-type stars and
early-type M dwarfs was investigated, e.g., by \citet{Pizzolato03} and
\citet{Reiners07}. They find that in slowly rotating stars, i.e.,
stars with $Ro = P/\tau_{\rm conv} \ga 0.1$, with $P$ the rotation
period and $\tau_{\rm conv}$ the convective overturn time, activity is
stronger at more rapid rotation. Above this threshold, activity
becomes saturated and does not grow as the stars rotate more rapidly.
For early- and mid-type M stars (M2--M6), \citet{Reiners09} showed
that magnetic fields also saturate at this threshold.

Among ultracool dwarfs, the existence of a rotation-activity
connection was probed by \citet{Reid02}, \citet{Mohanty03}, and
\citet{Reiners07}. While \citet{Reid02} report no relation between
rotation and activity, \citet{Mohanty03} showed a rotation-activity
connection down to spectral type M8 that breaks down at spectral type
M9. They also report an increase of the saturation velocity at mid-M
spectral types. \citet{RB07} report the first direct measurements of
photospheric magnetic fields from Zeeman line broadening, but their
sample is too small for a detailed investigation of the relation
between magnetic field strength, temperature, and rotation.

The temperature range of ultracool dwarfs is particularly interesting
for a number of reasons. Depending on age, stars and brown dwarfs
share the same temperature range so that young ultracool dwarfs can be
brown dwarfs while these have evolved to later spectral type at higher
ages. Thus, old ultracool dwarfs are always stars. It is also this
temperature range where the atmospheric conditions affecting the
high-energy processes of stellar activity are rapidly changing. At
lower temperature, the atmosphere becomes more and more neutral, which
can make a coupling between magnetic field lines and the atmosphere
more difficult \citep{Meyer99, Mohanty02}.

In order to investigate the physics of ultracool dwarfs, we observed a
large sample of objects with spectral types between M7 and M9.5. We
have introduced the sample in a first paper \citep[][Paper~I in the
following]{I}, where we searched for Li absorption lines and derived
space motion and the kinematic velocity of the sample. In this paper,
we focus on rotation, activity, and magnetic field measurements.

\section{Sample selection and observations}

\subsection{The sample}

We constructed a volume-limited sample ($d < 20$\,pc) of known M stars
of spectral type M7--M9.5. The targets are taken from several
catalogues and discoveries from the DENIS survey \citep{Delfosse01,
  Crifo05, PhanBaoBessel, PhanBao06} and the 2MASS survey
\citep{Cruz07}.

From the two surveys, we constructed a joint sample of objects within
20\,pc (for a more detailed description, see paper I).  Both surveys
together cover more than 50\% of the sky and can be considered as
almost complete in the spectral range M7--M9.5.  The full
``20pc-2MASS-DENIS'' survey contains 63 objects, 4 of which are found
in both the 2MASS and the DENIS surveys.

\subsection{Observations}

Observations for our 63 sample targets were collected using the HIRES
spectrograph at Keck observatory for targets in the northern
hemisphere, and using UVES at the Very Large Telescope at Paranal
observatory for targets in the southern hemisphere (PIDs 080.D-0140
and 081.D-0190).  HIRES data were taken with a 1.15\arcsec\ slit
providing a resolving power of $R \approx 31,000$. The three HIRES
CCDs cover the spectral range from 570 to 1000\,nm in one exposure.
UVES observations were carried out using a setup centered at 830\,nm
covering the spectral range 6400--10,200\,nm with a 1.2\arcsec\ slit
($R \approx 32,000$).  All data provide the H$\alpha$ line as well as
the absorption bands of molecular FeH around 1\,$\mu$m that are
particularly useful for the determination of rotation and magnetic
fields in M dwarfs \citep[see][]{Reiners06}.

Data reduction followed standard procedures including bias
subtraction, 2D flat-fielding, and wavelength calibration using ThAr
frames. HIRES data was reduced using the MIDAS echelle environment.
For the UVES frames, we used the ESO CPL pipeline, version 4.3.0. We
refer to Paper~I for more details.

\section{Measurements}

\subsection{H$\alpha$ Emission}

Our measurement of H$\alpha$ emission follows the procedure outlined
in \citet{Reiners08}.  To measure the equivalent width in the
H$\alpha$ line against the continuum, we normalize the line at two
footpoints blue- and redward of H$\alpha$. The footpoints are the
median values at 6545 -- 6555\,\AA\ on the left hand side, and at 6572
-- 6580\,\AA\ on the right hand side of the H$\alpha$ line.  None of
the emission lines found in our targets extends into the region used
for normalization.  The H$\alpha$ equivalent width is then measured by
integrating the flux from 6555 to 6572\,\AA. Equivalent widths of
H$\alpha$ emission lines are reported together with results on
rotation and magnetic fields in Table\,\ref{tab:results}. The typical
uncertainty is about a fifth of an {\AA}ngstr\"om. We convert the
measured H$\alpha$ equivalent width into H$\alpha$ flux, $F_{{\rm
    H}\alpha}$, by multiplying it with the flux per unit equivalent
width from the continuum flux in synthetic spectra. We divide the
H$\alpha$ flux by the bolometric flux, $F_{\rm bol} = \sigma T^4$, and
use $F_{{\rm H}\alpha}/F_{\rm bol} = L_{{\rm H}\alpha} / L_{\rm bol}$
to derive the normalized $H\alpha$ luminosity
\citep[see][]{Reiners08}. To find the temperature of our targets, we
used the spectral types and calculated $T_{\rm{eff}}$ according to the
conversion given in \cite{Golimowski04}.

\subsection{Rotation and Average Magnetic Field}

The analysis of our spectra follows the strategy laid out in
\citet{Reiners06, RB07} and \citet{Reiners09}. To measure the
projected rotation velocity $v\,\sin{i}$ and the magnetic field
average over the visible surface, $Bf$ (with $f \le 1$), we utilize
the absorption band of molecular FeH close to 1\,$\mu$m. We compare
our data to spectra of the slowly rotating M-stars GJ\,1002 (M5.5) and
Gl\,873 (M3.5). The average magnetic flux of Gl\,873 was measured to
be $Bf = 3.9$\,kG using an atomic FeI line \citep{JKV00}. In order to
match the absorption strength of the target spectra, the intensity of
the FeH absorption lines in the two comparison spectra is adjusted
according to an optical-depth scaling \citep[see][]{Reiners06}; the
FeH intensity is a free parameter in our fit. Note that in the
magnetic field measurement of our template star, Gl\,873, $Bf$ is the
weighted sum of several magnetic components used for the fit ($\Sigma
Bf$). In our data, we cannot follow such an approach because (1) the
spectral resolution of our data does not allow the differentiation of
individual magnetic components, and (2) no information on the magnetic
splitting pattern of FeH lines is used.  Thus, with our method we can
only scale the product $Bf$ from the spectrum of Gl\,873. For a
detailed discussion of the systematic uncertainties, we refer to
\citet{Reiners06}. The typical uncertainty in $v\,\sin{i}$ is
$\sim$2\,km\,s$^{-1}$ for the slow rotators and $\sim 10$\,\% in the
case of rapid rotation. The uncertainty in $Bf$ is usually several
hundred Gauss.

\begin{figure}
    \parbox{0.45\hsize}{
      \centering
      \mbox{\includegraphics[width=\hsize]{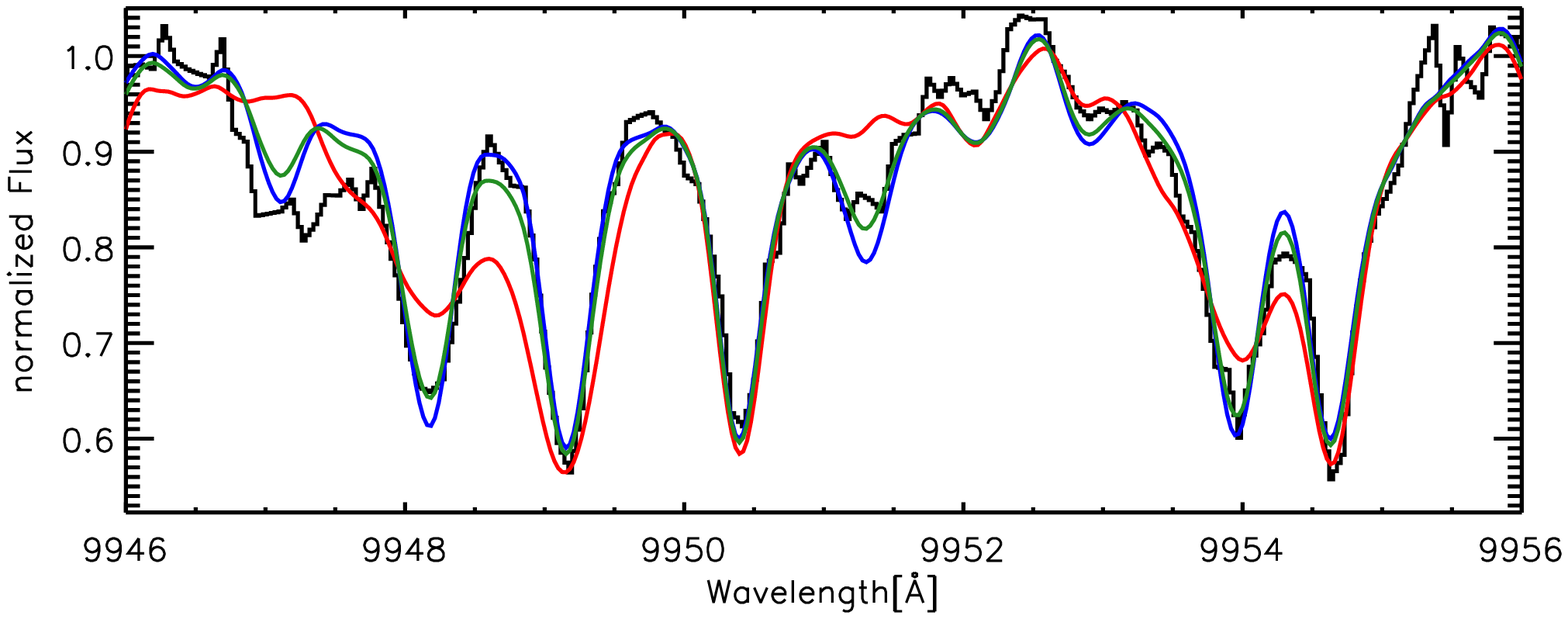}}\\
      \mbox{\includegraphics[width=.97\hsize]{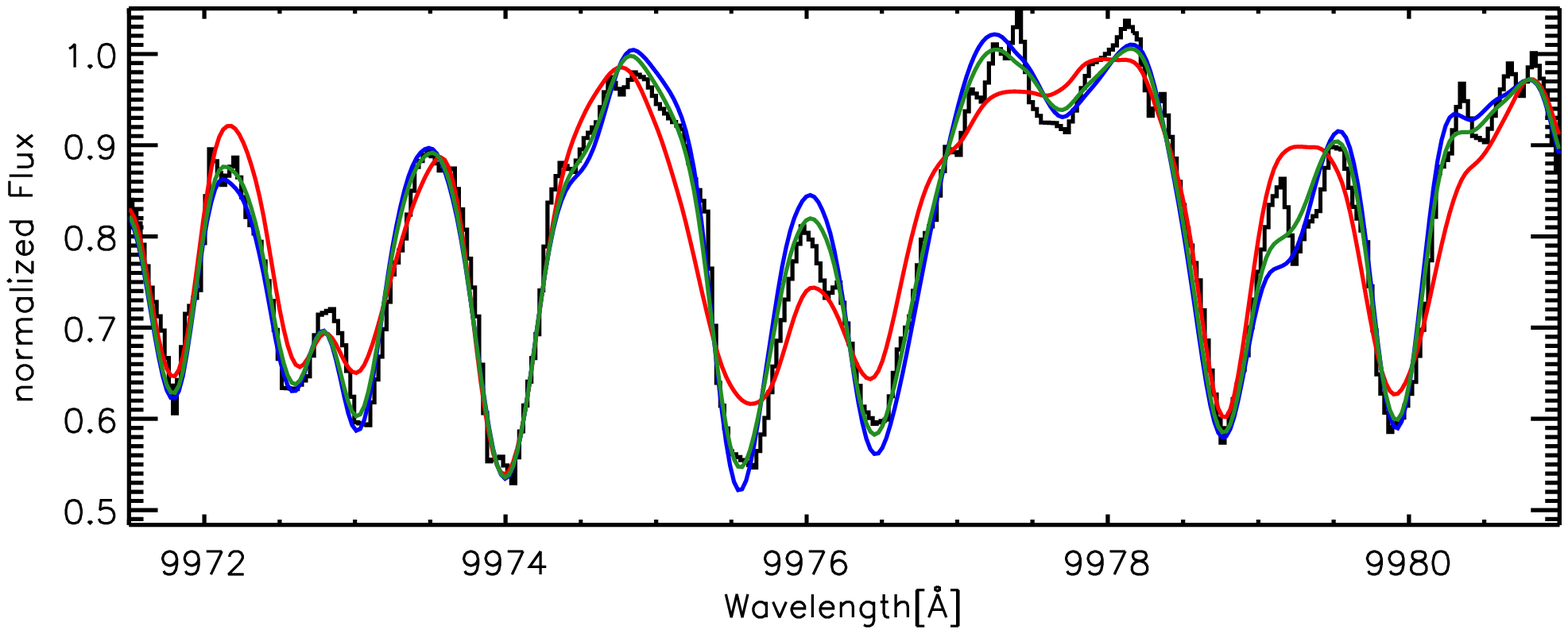}}\\[2mm]
      \mbox{\includegraphics[width=.96\hsize]{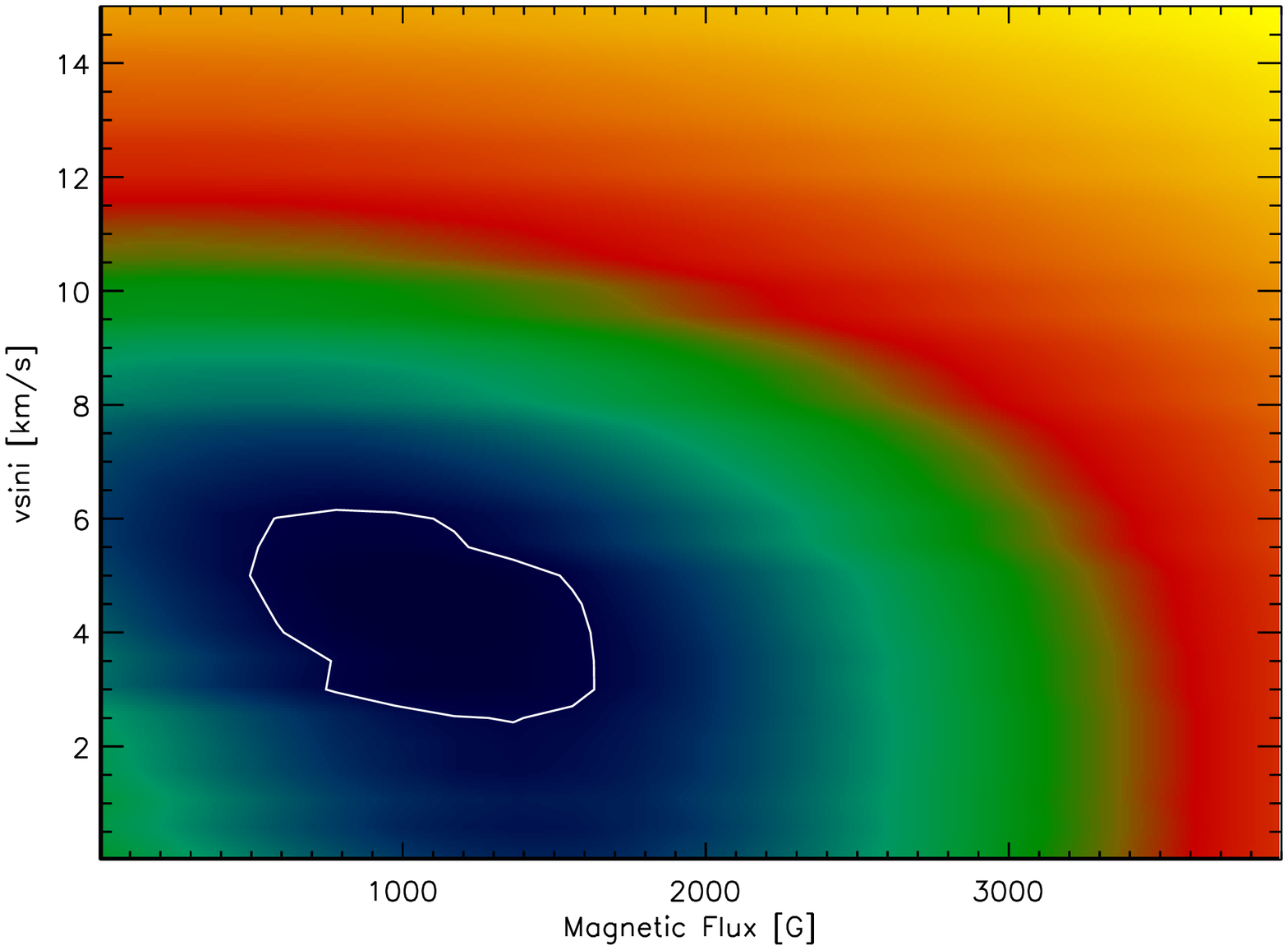}}}
    \caption{\label{fig:chisqexample}Data and fit (top panel) and
      $\chi^2$-landscapes (bottom panel) for 2MASS~0818580+233352. In
      the upper panel, data are shown as black histograms. The three
      coloured lines show our fit for no magnetic field (blue line),
      strong magnetic flux ($Bf \sim 4$\,kG, red line), and the best
      fit, which is an interpolation of the two (green line). In the
      bottom panel, dark and blue colour indicates low $\chi^2$
      values, red and yellow colours show bad fit quality (high values
      of $\chi^2$. The white line shows the 3$\sigma$-level.}
\end{figure}

For the determination of the average magnetic field, $Bf$, we
concentrate on relatively small wavelength regions that contain
absorption lines particularly useful for this purpose, i.e., regions
that contain some magnetically sensitive as well as magnetically
insensitive lines.  We determine $Bf$ of our target stars by
comparison of the spectral regions at 9946.0--9956.0\,\AA\ and
9972.0--9981.0\,\AA. We show an example of the data and the fit
quality together with a $\chi^2$ map in $v\,\sin{i}$ and $Bf$ in
Fig.\,\ref{fig:chisqexample}.  In the upper two panels, we plot the
data and two extreme cases (a) without any magnetic field, and (b)
strong magnetic flux with $Bf \approx 4$\,kG. Template spectra are
artificially broadened and scaled to match the absorption depth of the
FeH lines in our sample targets. The lower panel of
Fig.\ref{fig:chisqexample} shows a $\chi^2$-map, i.e., the goodness of
fit, $\chi^2$, as a function of projected rotational velocity,
$v\,\sin{i}$, and average magnetic field, $Bf$. We mark the formal
3$\sigma$-region for two free parameters, $\chi^2 = \chi^2_{\rm min} +
11.8$, with a white line \citep{Press92}. To determine our
3$\sigma$-uncertainties in $v\,\sin{i}$ and $Bf$, we are marginalizing
over the other component so that we can chose the range $\chi^2 =
\chi^2_{\rm min} + 9$ because the other parameter was varying freely.
The values of the reduced $\chi^2$, $\chi^2_{\nu}$, are on the order
of unity except for the spectroscopic binary, 2MASS J$0435161-160657$,
for which no rotation and magnetic flux analysis could be carried out.
In one case, 2MASS J0320596+185423 (LP412-31), $Bf$ is probably larger
than the maximum limit of our method (3.9\,kG). Fit quality is
therefore not optimal in this case, but the differences are likely to
be explained by the effect of a stronger field \citep[see also Fig.~8
in][]{RB07}.

The results of our analysis for all stars are given in
Table\,\ref{tab:results}. Some of the stars were already analyzed in
\citet{RB07} but have been consistently reanalyzed following the
outlined strategy together with the new targets. Values of
$v\,\sin{i}$ and $Bf$ are best fits from our $\chi^2$ analysis with
formal 3$\sigma$-uncertainties.  The uncertainties in $Bf$ are usually
on the order of a few hundred Gauss. The typical uncertainty in
$v\,\sin{i}$ is on the order of 10\,\% with a minimum uncertainty of
2\,km\,s$^{-1}$. Our detection limit due to limited spectral resolving
power is $v\,\sin{i}_{\rm{min}} = 3$\,km\,s$^{-1}$
\citep[see][]{RB07}.

\section{Results}

\begin{figure}
  \plotone{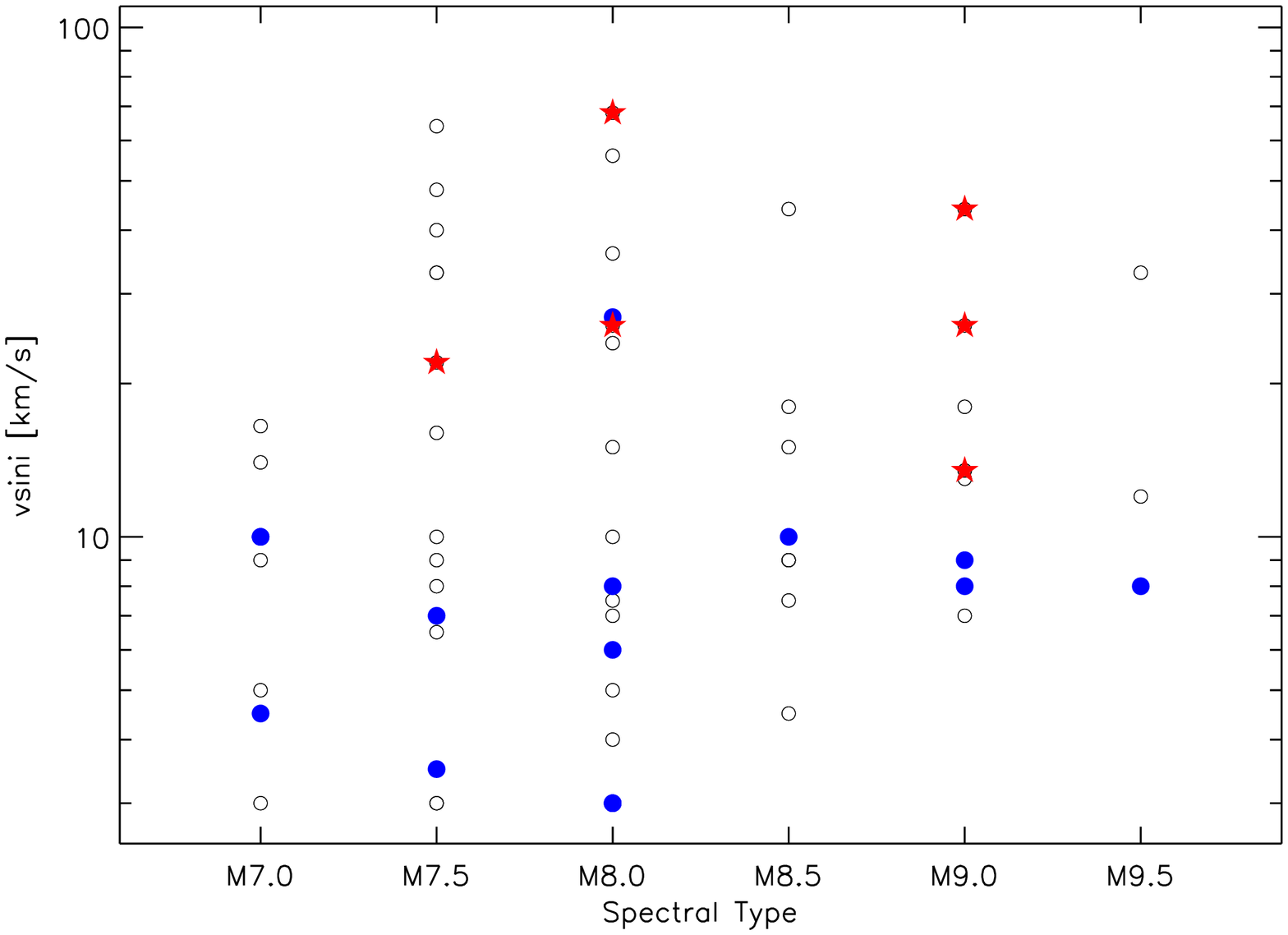}
  \caption{\label{fig:vsiniSpType}Projected rotation velocity
    $v\,\sin{i}$ as a function of spectral type. Blue filled circles
    are members of the old population, red stars are young brown
    dwarfs with Li detection. Note that $v\,\sin{i}$ is plotted on a
    log-scale, and that the old stars are predominatly occupying the
    region $v\,\sin{i} < 10$\,km\,s$^{-1}$.}
\end{figure}

\begin{figure}
  \plotone{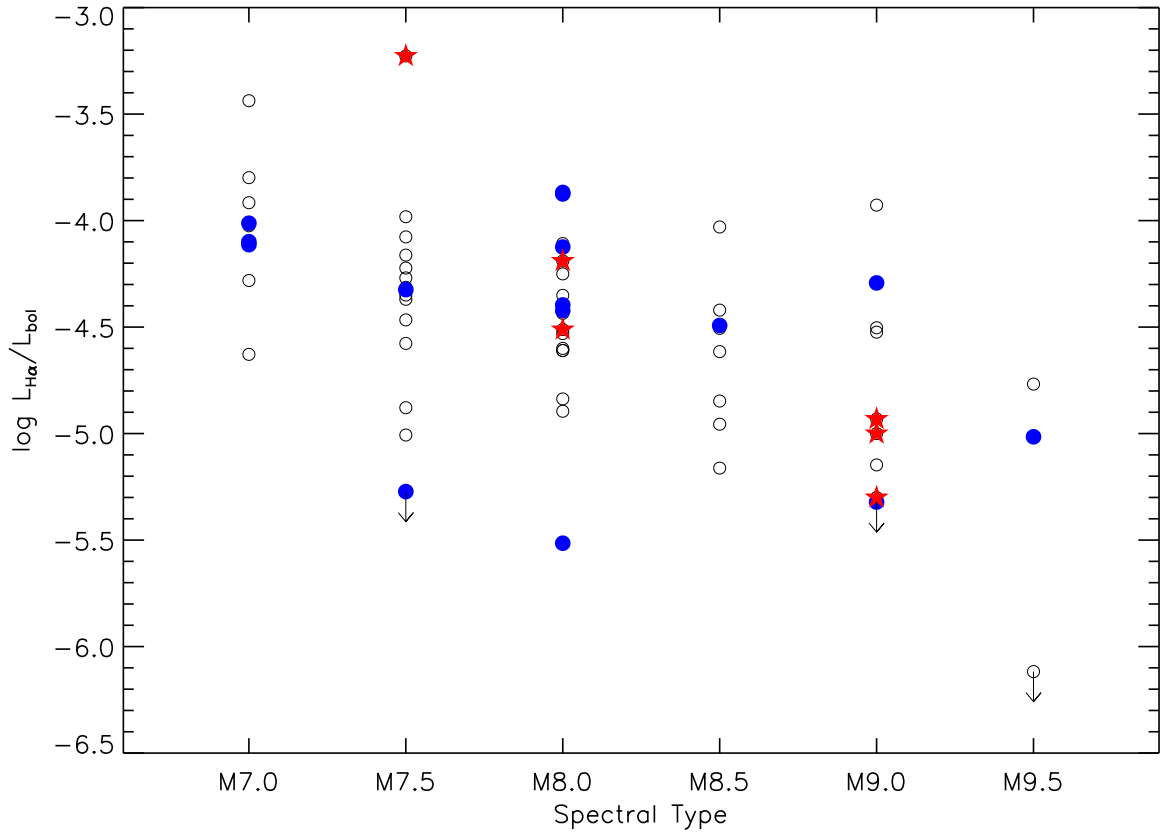}
  \caption{\label{fig:HalphaSpType}Normalized H$\alpha$ luminosity as
    a function of spectral type. Symbols as in
    Fig.\,\ref{fig:vsiniSpType}.}
\end{figure}

Projected rotational velocity, $v\,\sin{i}$, normalized H$\alpha$
luminosity, $\log{L_{\rm H\alpha}/L_{\rm bol}}$, and average magnetic
field, $Bf$, are plotted as a function of spectral type in
Figs.\,\ref{fig:vsiniSpType}, \ref{fig:HalphaSpType}, and
\ref{fig:BfSpType}, and are discussed in the following. Li brown
dwarfs as well as stars with a space velocity component $V <
-30$\,km\,s$^{-1}$ are indicated (see Paper~I). The latter sample
probably consists of predominantly relatively old objects, although
this is not necessarily true for each individual object
\citep{Wielen77}.

\subsection{Rotation}

We find rotation velocities up to $v\,\sin{i} \approx
70$\,km\,s$^{-1}$ in our sample. The Li brown dwarfs exhibit
relatively high rotation velocities (with a mean of $v\,\sin{i} =
33$\,km\,s$^{-1}$) while rotation in the old sample is very low (mean
$v\,\sin{i} = 8$\,km\,s$^{-1}$). This is a clear indication that
rotational braking slows down stars of spectral type M7--M9.5 as they
age.  Nevertheless, the stars of the old sample also exhibit
significant rotation, which stands in contrast to old early-M type
dwarfs.  The longer timescales of rotational braking appear very clear
in this sample \citep[see][]{West08, Reiners08}.

\begin{figure}
  \plottwo{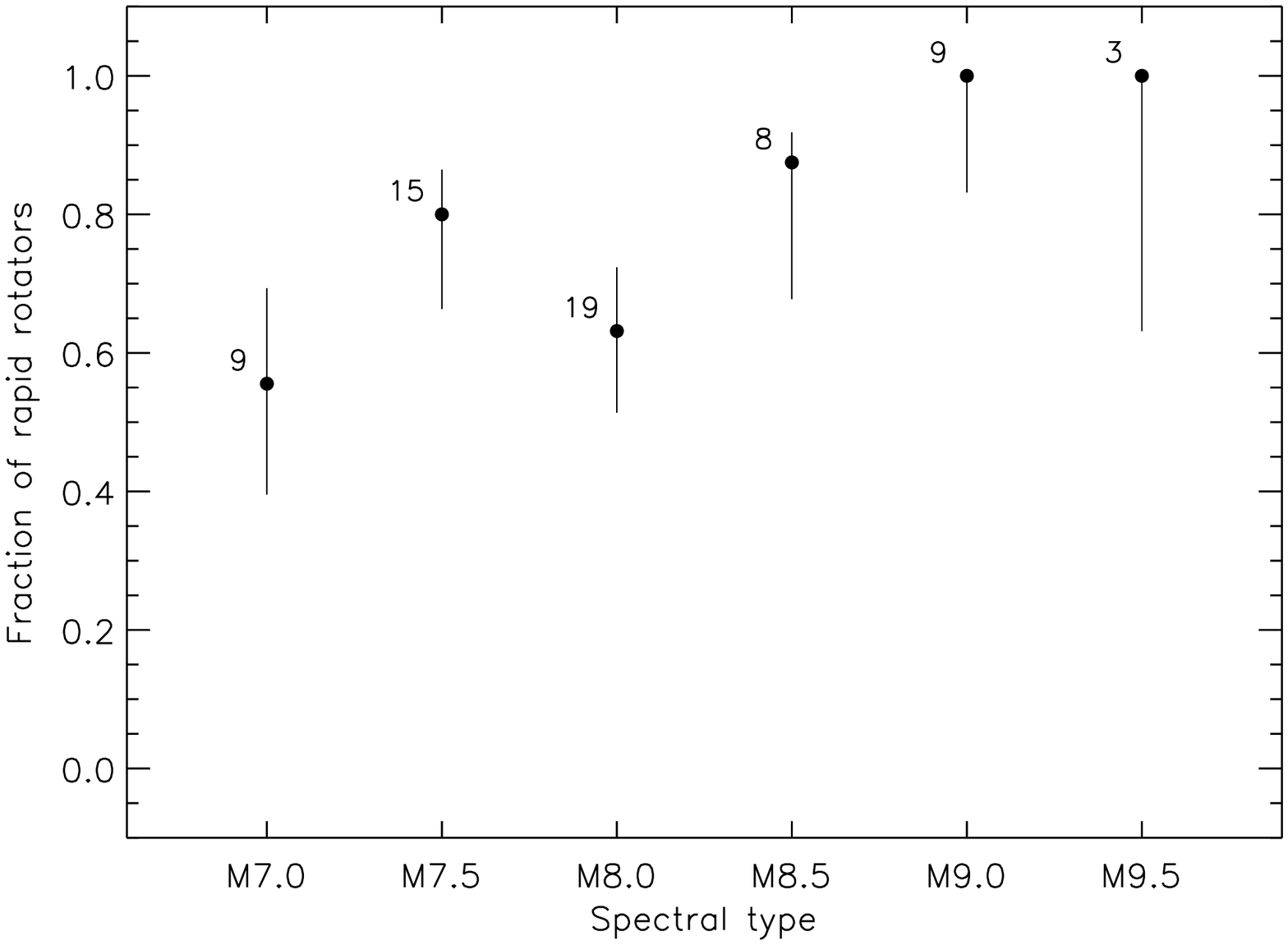}{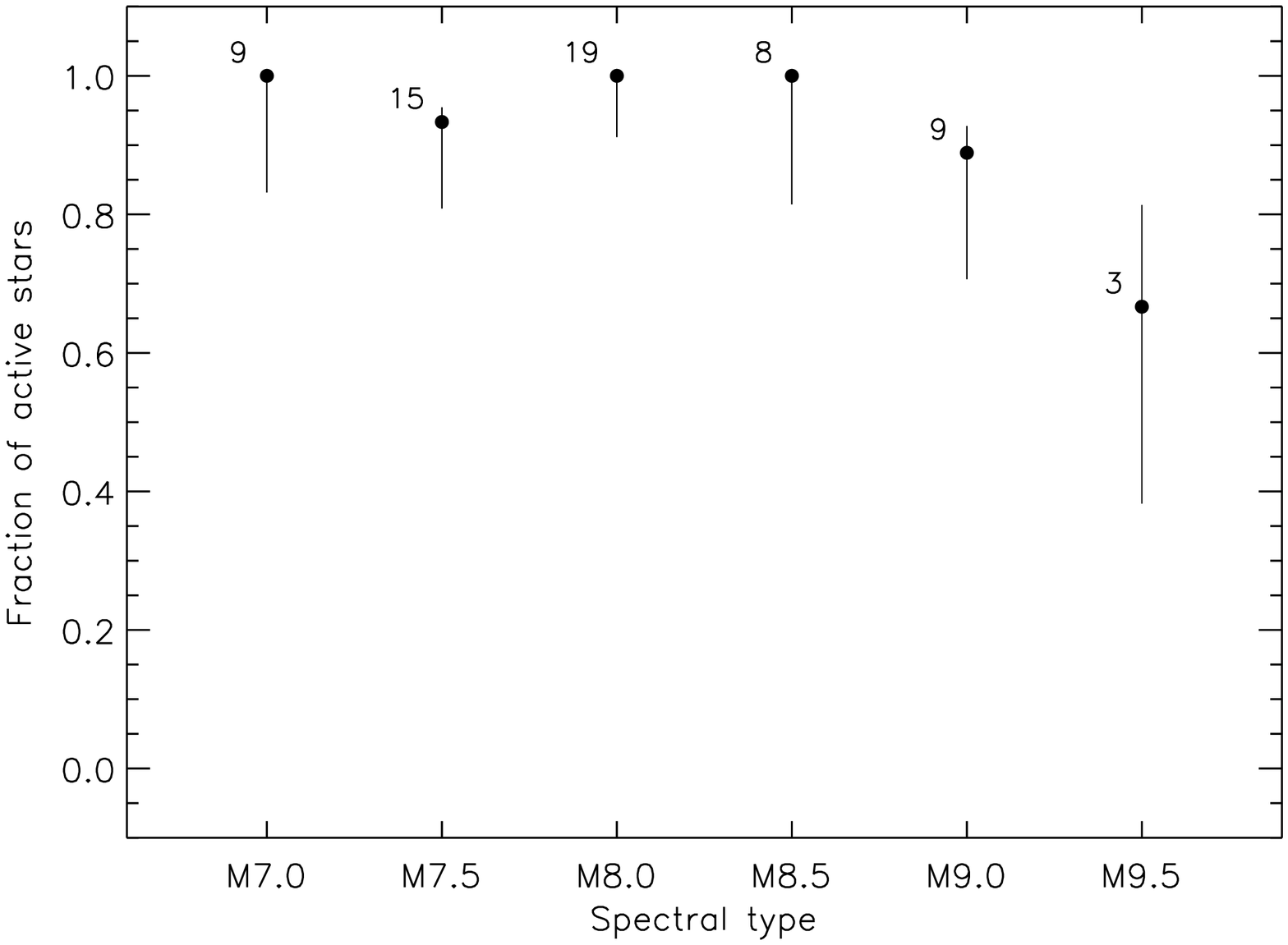}
  \caption{\label{fig:ActFrac}Fraction of stars rotating at 5\,km/s or
    more \emph{(left)} and fraction of active stars \emph{(right)}.
    Error bars show 1$\sigma$-uncertainties. }
\end{figure}

\begin{figure}
  \plotone{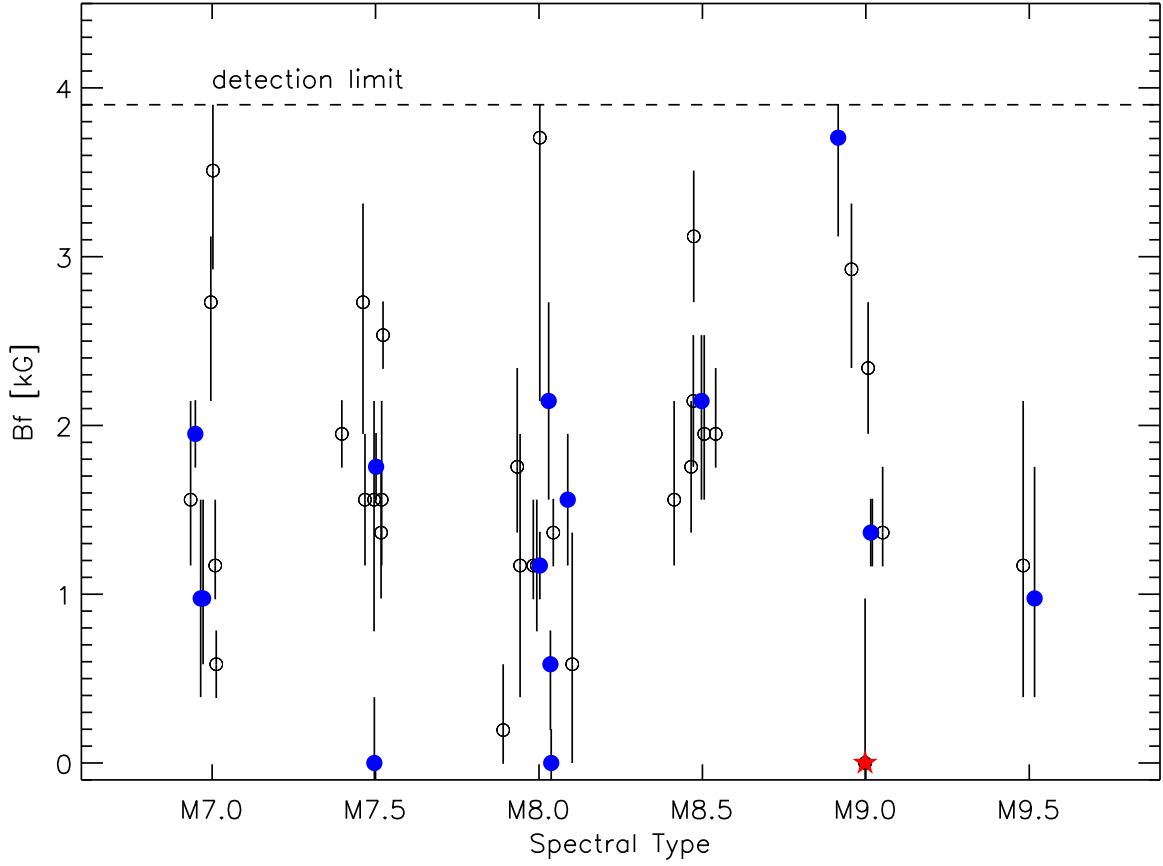}
  \caption{\label{fig:BfSpType}Average magnetic field, $Bf$ as a
    function of spectral type. Symbols as in
    Fig.\,\ref{fig:vsiniSpType}. Positions in spectral type are
    plotted with a small offset for clarity.}
\end{figure}

In the left panel of Fig.\,\ref{fig:ActFrac}, we show the fraction of
``rapid'' rotators, i.e. stars with $v\,\sin{i} > 5$\,km\,$^{-1}$.
The fraction of stars in our local sample populating this velocity
range is about 60\,\% at spectral type M7. At spectral class M9, we
find no object that is rotating slower than 5\,km\,s$^{-1}$.
Statistically, we cannot draw a robust conclusion from this limited
sample, i.e., we cannot reject the hypothesis that the fraction of
stars rotating more rapidly than $v\,\sin{i} > 5$\,km\,$^{-1}$ is
constant between M7 and M9.5. However, the trend that late-M dwarfs
seem to be generally rotating more rapidly than early- and mid-M
dwarfs is supported by other samples extending into the L-dwarf regime
\citep{Mohanty03, Reiners08}, and seems to be rather robust. Such a
trend is an important constraint for surveys searching for radial
velocity variations with the goal to find planets around low-mass
stars, because the achievable radial velocity accuracy strongly
depends on rotational broadening.

\subsection{H$\alpha$-Activity}

While rotation of ultra-cool dwarfs is braked with time, no age effect
is visible in chromospheric H$\alpha$ activity
(Fig.\,\ref{fig:HalphaSpType}). It is known that the mean level of
normalized H$\alpha$ activity is diminishing from mid-M to late-M and
L spectral classes. In our sample, we observe a mean level of
$\log{L_{\rm H\alpha}/L_{\rm bol}} = -4.2$ at spectral type M7, and of
$\log{L_{\rm H\alpha}/L_{\rm bol}} = -4.8$ at spectral type M9.
Interestingly, we see no clear dependence of normalized H$\alpha$ on
age; we find the few Li brown dwarfs to be on the average level of
H$\alpha$ luminosity at spectral type M8 and at the lower activity
range at M9. Note that the high H$\alpha$ emission in the Li brown
dwarf at M7.5 is probably due to accretion.

In the right panel of Fig.\,\ref{fig:ActFrac}, we plot the fraction of
active stars among our sample (with 1$\sigma$-uncertainties). Among
the 63 objects, we found only three objects without any sign of
H$\alpha$ emission. All other 60 objects exhibit H$\alpha$ in
emission. The activity fraction in our sample is close to 100\,\% at
spectral types M7--M8.5 ($98^{+1}_{-6}\,\%$, only one out of 41 shows
no H$\alpha$ emission). On the other hand, we find a fair fraction of
stars with no H$\alpha$ emission at spectral class M9--M9.5
($83^{+7}_{-15}\,\%$). It is important to realize that the
non-detection of H$\alpha$ in our data does not exclude the presence
of activity altogether; at some low level, all stars might show
H$\alpha$ emission. Again, from our sample it is not possible not draw
robust statistical conclusions about a dependence of activity on
spectral type, but a turnover of the activity fraction at spectral
type M9 \citep[cp.][]{West08} seems to be very plausible.

\subsection{Average Magnetic Field}

In the magnetically sensitive absorption lines of molecular FeH, we
can measure average magnetic fields in relatively rapidly rotating M
dwarfs. We report magnetic field measurements for all stars of our
sample with a projected rotation velocity of $v\,\sin{i} <
20$\,km\,s$^{-1}$. Above that rotation rate, spectral lines are too
broad for a reliable detection of Zeeman broadening. In all 46 stars
that fulfill this criterion, our 3\,$\sigma$ limits allow a
determination of $Bf$ within an uncertainty less or equal to 1\,kG,
with a much smaller uncertainty in the slow rotators.

We detect non-zero magnetic fields in 41 of our 46 targets, i.e., only
in 5 (11\,\%) of our objects, the average magnetic field is on the
order of only a few hundred Gauss or less. Our measurements of $Bf$
are plotted as a function of spectral type in
Fig.\,\ref{fig:BfSpType}. At all spectral types, the scatter among
average magnetic field values is quite large. We find both, weak and
strong fields in M7, M8, and M9 objects. The mean average magnetic
field is $Bf \approx 1600$\,G with an rms of 900\,G. The mean of our
3$\sigma$ uncertainties is $\sim 430$\,G implying that the scatter in
$Bf$ is much larger than the formal statistical uncertainties. There
is some evidence for a lack of very weak fields among the M8.5--M9.5
stars; all 15 stars in this spectral range show average fields on the
order of 1~kG or stronger. Interestingly, the only object with a very
weak field is the Li brown dwarf 2MASS~J0443376+000205. On the other
hand, six out of 30 stars with spectral types M7.0--M8.0 have average
magnetic fields weaker than 900\,G. Performing a K-S test on the two
distributions among the stars only, we find that the hypothesis of a
single underlying distribution for the average magnetic fields in the
M7--M8 and M8.5--M9.5 spectral bins can only be rejected at the
1$\sigma$-level (and the level is below 1$\sigma$ if we include the
brown dwarf).

\cite{RBC09} found that accreting young brown dwarfs seem to have very
weak fields, much weaker than accreting stars at the same age (but
earlier spectral type). There is mounting evidence that it is
difficult to produce magnetic fields of kilo-Gauss strength at young
ages in brown dwarfs.  Whether this applies to brown dwarfs in
general, or is due to the young age of very-low mass objects, is not
yet clear and can only be decided from magnetic field observations of
older brown dwarfs, which are currently not available.

Other than this, we see no evidence for a dependence between spectral
type and average magnetic field, in particular no rule for the
production of very strong fields. Furthermore, there is no clear
evolution of magnetic field strength with age.

\subsection{The Connection between Rotation, Activity, and Magnetic
  Flux}
\label{sect:rotact}

Solar-type stars as well as early-M stars show a well-documented
relation between rotation and magnetic activity
\citep[e.g.,][]{Pizzolato03, Reiners07, Reiners09}. This
rotation-activity relation is generally explained by stellar activity
being proportional to magnetic flux generation. The latter is ruled by
the Rossby number, $Ro = P/\tau_{\rm conv}$, the ratio between the
rotation period and the convective overturn time.  For $Ro > 0.1$,
magnetic flux generation and activity are proportional to the stellar
rotation rate. Below that threshold, activity and magnetic field
generation are saturated and do not grow with decreasing $Ro$.

Mainly because of the difference in radius between sun-like and M-type
stars, the threshold surface rotation velocity is a few ten
km\,s$^{-1}$ in sun-like stars, but only on the order of
1\,km\,s$^{-1}$ in mid- and late-M dwarfs. Thus, essentially all M
dwarfs with detectable rotation (typically $v\,\sin{i} \ga
3$\,km\,s$^{-1}$ for most spectrographs) are in the saturated regime
\citep[compare][]{Reiners07}. The relation between Rossby number and
activity has been expected to break down in fully convective stars,
because the main motivation for the Rossby dependence comes from the
assumption of an $\alpha\Omega$-dynamo that is probably only operating
in the presence of a radiative core and not in fully convective stars
\citep{Durney93}.  It is therefore assumed that in mid- to late-M
dwarfs, the rotation-activity relation is no longer valid or changes
substantially.

Three high-resolution surveys investigated the rotation-activity
relation beyond the threshold to complete convection, i.e., in objects
of spectral type mid-M and later. \citet{Reid02} find no connection
between rotation and activity in their sample of M7--M9.5 stars.
\citet{Mohanty03} show that a rotation-activity connection exists down
to spectral type M8.5, and that the connection is weaker at M9 until
it breaks down altogether among the L dwarfs. This result is supported
by \citet{Reiners08} who show that a $\Gamma$-shaped rotation-activity
relation exists in stars earlier than spectral type M9. Their sample
combine the objects from \citet{Mohanty03} and \citet{Reiners08}
showing that no relation between rotation and activity appears at
spectral types M9 and later. \citet{Mohanty03} add that there appears
to be a change in the magnitude of the threshold velocity above which
$L_{\rm H\alpha}/L_{\rm bol}$ saturates. While this velocity is about
3\,km\,s$^{-1}$ at early-M, it appears to be roughly 10\,km\,s$^{-1}$
at spectral types late-M according to that work.

In our new volume-limited sample of M7--M9.5 dwarfs, we can now
investigate the transition from the regime of an intact
rotation-activity relation towards the regime where this relation no
longer applies. Furthermore, we can look for a change in the threshold
velocity above which activity saturates.

We divide our sample of 62 dwarfs with $v\,\sin{i}$ measurements into
three subsamples of consecutive spectral types. To obtain samples of
approximately same size, we divide the samples into objects of
spectral type M7--M7.5 (23 stars), M8.0 (19 stars), and M8.5--M9.5 (20
stars).  For simplicity, we refer to these subsamples as M7, M8, and
M9, respectively. Normalized H$\alpha$ activity and average magnetic
field for each subsample are plotted as a function of projected
rotation velocity in Fig.\,\ref{fig:ActRot}. As a quantitative measure
for the connection between rotation, H$\alpha$-activity, and average
magnetic field, we calculate the linear correlation coefficients and
the Spearman rank correlation \citep[][, Chapter 14.6]{Press92}
between these parameters for the three subsamples. The coefficients
are given in Table\,\ref{tab:correlation}. While the correlation
coefficients cannot present a statistically meaningful measure of the
significance of a correlation (particularly not in the case of our
small sample size), the Spearman rank test provides a statistically
more robust indicator. For the Spearman rank tests, we give the
rank-order correlation coefficient, $r_s$, in the first column, and
the associated probability of a non-correlation in the second column
of Table\,\ref{tab:correlation}.

\begin{deluxetable}{lccc}
  \tablecaption{\label{tab:correlation}Correlations between rotation,
    H$\alpha$-activity, and the average magnetic field $Bf$. }
  \tablewidth{0pt}
  \tablehead{Relation & \colhead{M7} & \colhead{M8} & \colhead{M9\tablenotemark{a}}}
  \startdata
  \multicolumn{4}{c}{Correlation coefficients}\\
  \noalign{\smallskip}
  \hline
  \noalign{\smallskip}
  $\log{L_{\rm H\alpha}/L_{\rm bol}}$ vs. $v\,\sin{i}$ & 0.63 & 0.32 & $-$0.34 ($-$0.13)\\
  $Bf$ vs. $v\,\sin{i}$                                & 0.32 & 0.69 & $-$0.09 ($-$0.09)\\
  $\log{L_{\rm H\alpha}/L_{\rm bol}}$ vs. $Bf$         & 0.77 & 0.66 & \phantom{$-$}0.61 \phantom{$-$}(0.61)\\
  \cutinhead{Spearman rank test\tablenotemark{b}}
  $\log{L_{\rm H\alpha}/L_{\rm bol}}$ vs. $v\,\sin{i}$ & 0.62 & 0.17 & $-$0.13\, (0.01)\\
                                                       &0.006 & 0.52 & \phantom{$-$}0.62\, (0.97)\\
  $Bf$ vs. $v\,\sin{i}$                                & 0.32 & 0.41 & \phantom{$-$}0.06\, (0.06)\\
                                                       & 0.21 & 0.16 & \phantom{$-$}0.84\, (0.84)\\
  $\log{L_{\rm H\alpha}/L_{\rm bol}}$ vs. $Bf$         & 0.58 & 0.56 & \phantom{$-$}0.56\, (0.56)\\
                                                       & 0.01 & 0.05 & \phantom{$-$}0.02\, (0.02)  
 \enddata
 \tablenotetext{a}{Numbers in parentheses are excluding M9 objects
   faster than $v\,\sin{i} = 18$\,km\,s$^{-1}$ (see text).}
 \tablenotetext{b}{The upper row always gives the rank-order
   correlation coefficient, $r_s$. The lower row gives the probability
   of a non-correlation.}
\end{deluxetable}

\subsubsection{Rotation and H$\alpha$-Activity}

Evidence exists from different earlier works that the
rotation-activity relation undergoes a change around late-M spectral
types. The new sample presented here allows for the first time to
study this relation in detail in the narrow range M7--M9.5 and in a
substantial number of objects that occupy a wide range in rotation
velocity.

\begin{figure}
  \mbox{
    \mbox{\includegraphics[width=.3\hsize]{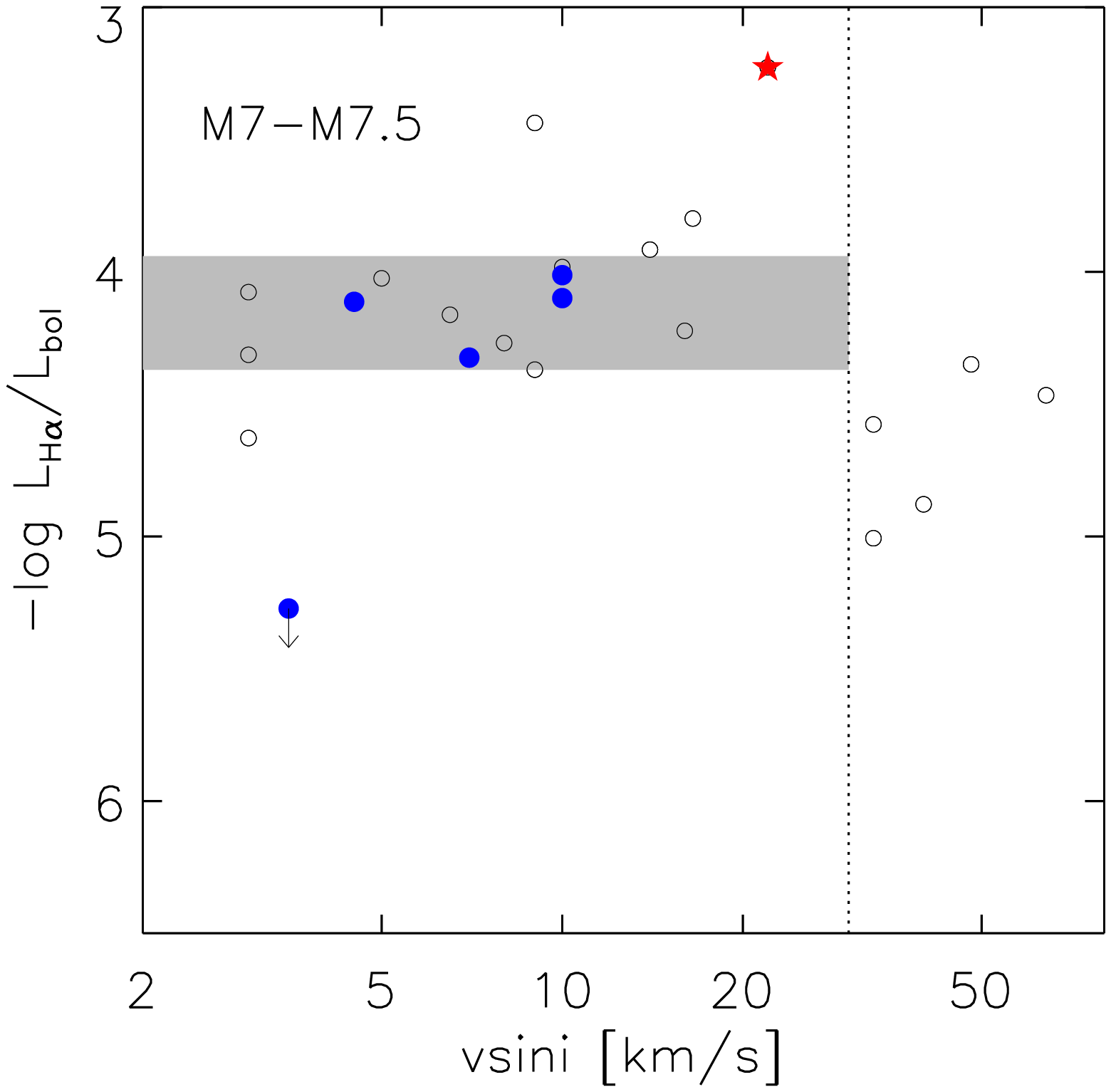}}
    \mbox{\includegraphics[width=.3\hsize]{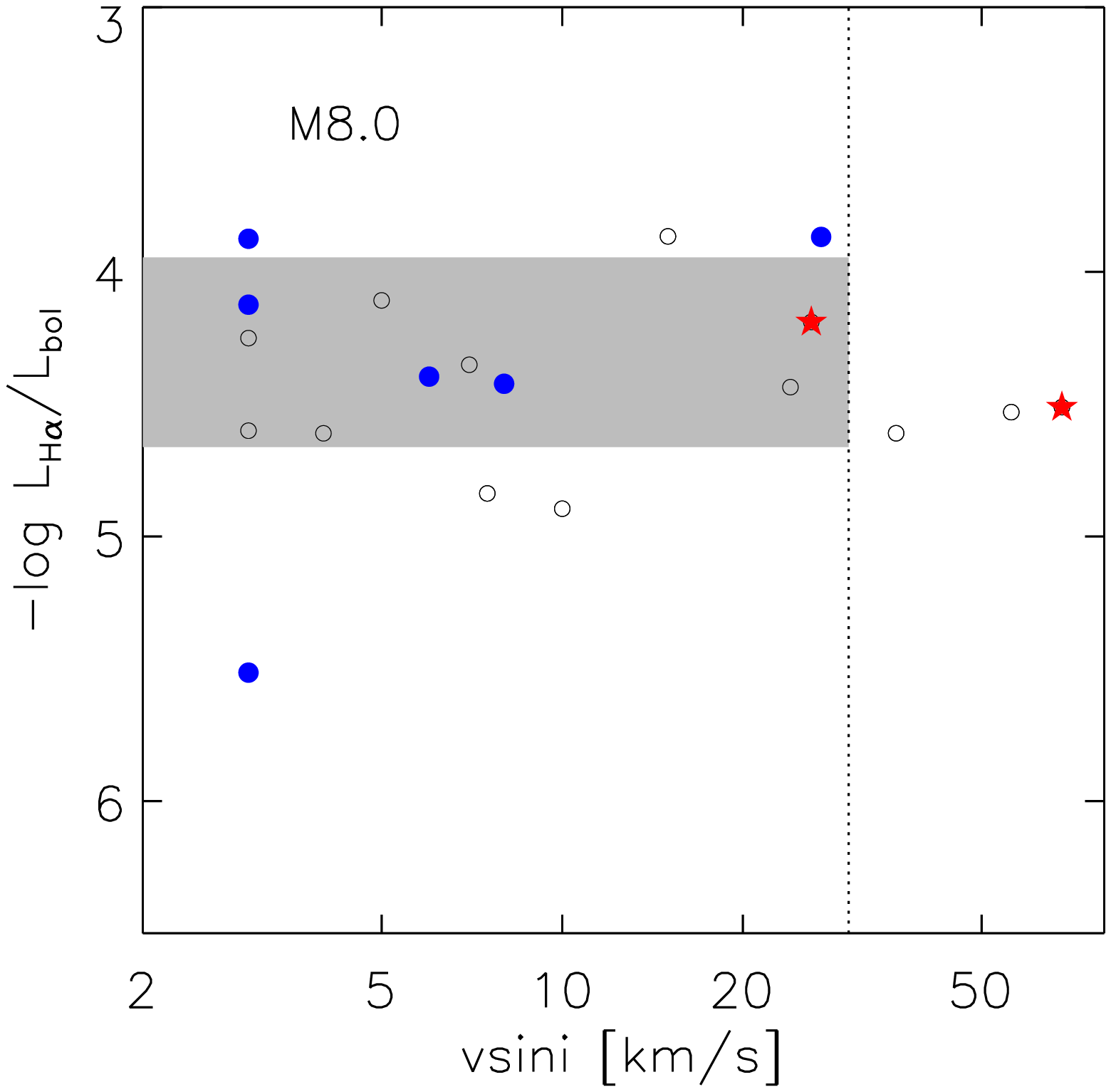}}
    \mbox{\includegraphics[width=.3\hsize]{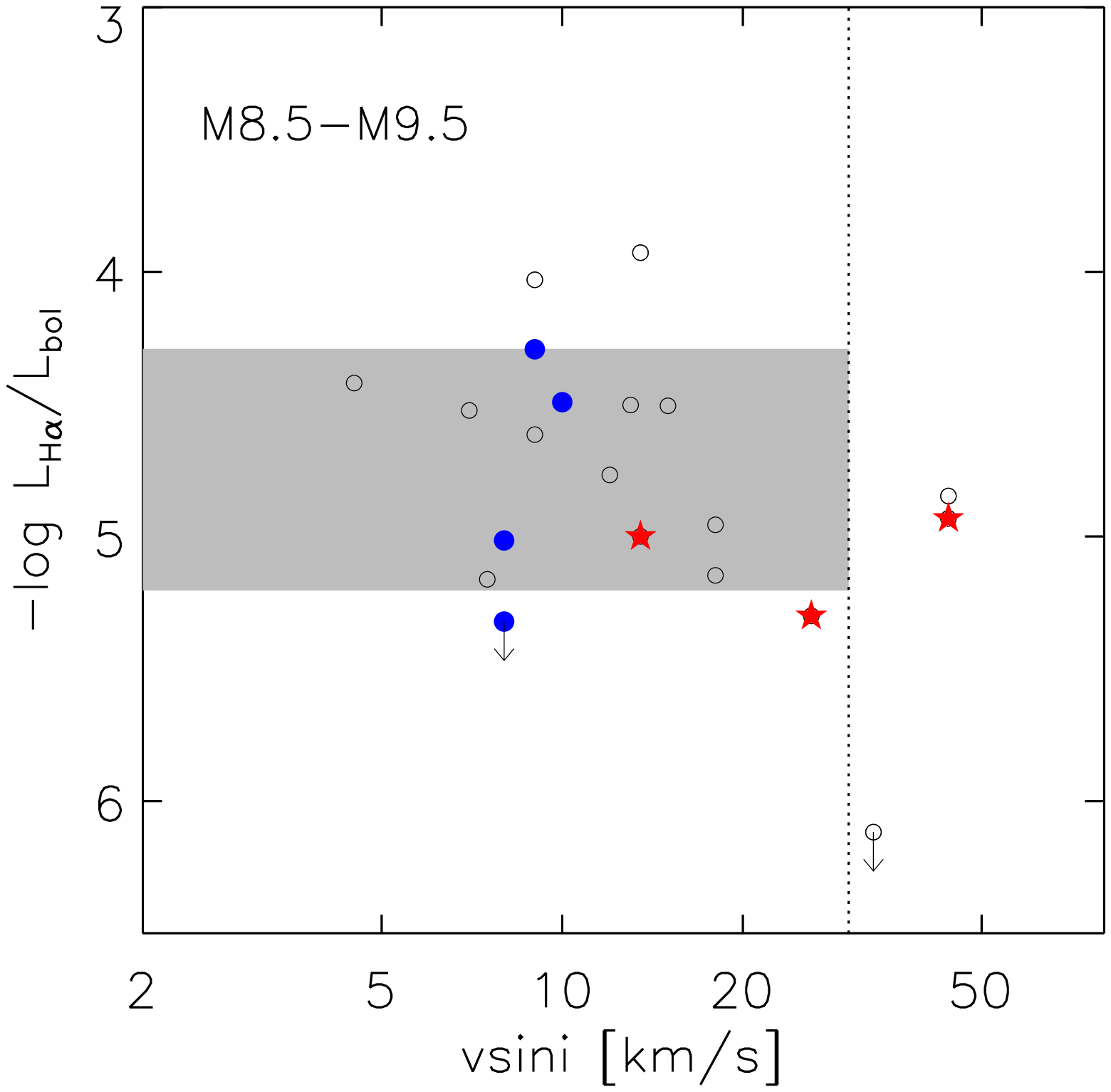}}
  }\\
  \mbox{
    \mbox{\includegraphics[width=.3\hsize]{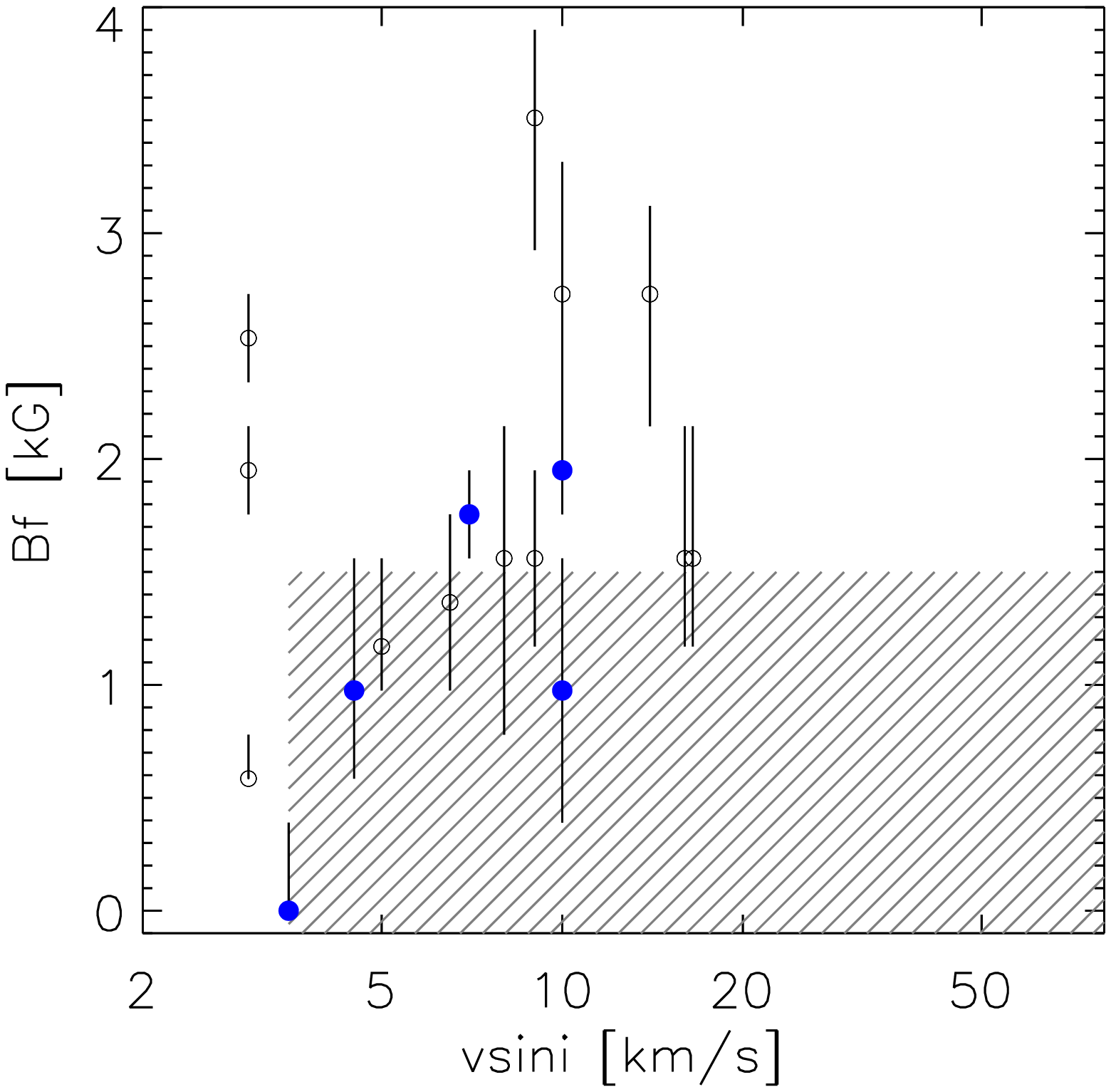}}
    \mbox{\includegraphics[width=.3\hsize]{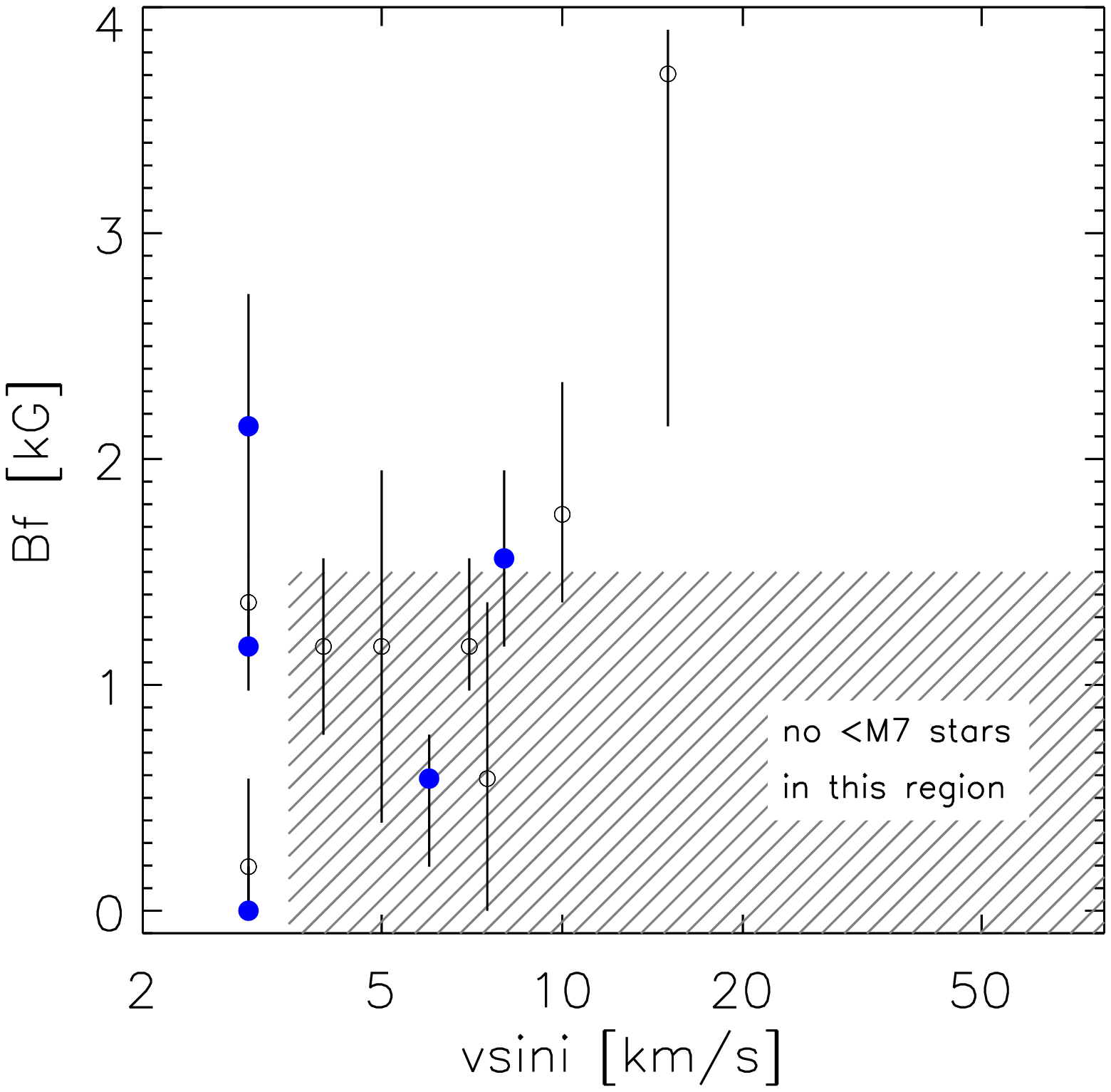}}
    \mbox{\includegraphics[width=.3\hsize]{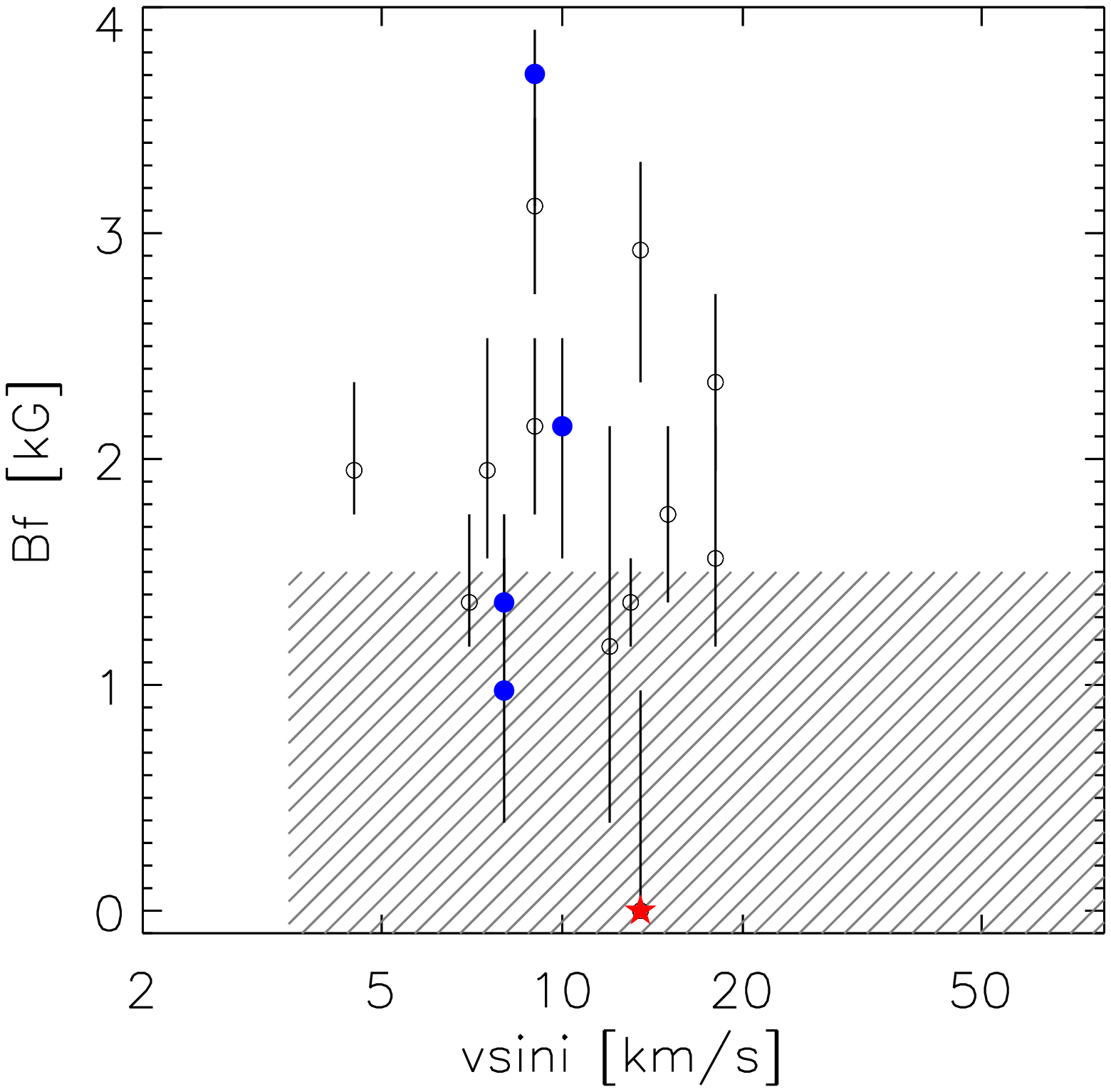}}
  }
  \caption{\label{fig:ActRot}Normalized H$\alpha$ luminosity and
    average magnetic field as a function of rotation velocity for
    different spectral types. Left, center, and right panels show
    spectral types M7--M7.5, M8.0, and M8.5--M9.5, respectively.
    \emph{Top panel:} Normalized H$\alpha$ luminosity as a function of
    spectral type. The grey shaded areas visualize the
    1$\sigma$-region around the mean for objects with $v\,\sin{i} <
    30$\,km\,s$^{-1}$ (see Fig.\,\ref{fig:HistLHalpha}).  \emph{Bottom
      panel:} Average magnetic field, $Bf$, as a function of rotation
    velocity. No measurement of $Bf$ is possible at $v\,\sin{i} >
    20$\,km\,s$^{-1}$. The hatched region shows the area that is not
    occupied by stars earlier than M7 \citep[see Fig.\,5
    in][]{Reiners09}. In all panels, red stars indicate young brown
    dwarfs with Li, blue filled circles show members of the old
    population as in Figs.\,\ref{fig:vsiniSpType}--\ref{fig:BfSpType}.
    The strong H$\alpha$ emission of the very active, young M7 object
    is probably due to accretion.}
\end{figure}

From the upper panel of Fig.\,\ref{fig:ActRot}, the first thing we
note is that there is evidence for an effect of supersaturation.  In
the M7 sample, objects with $v\,\sin{i} > 30$\,km\,s$^{-1}$ have
smaller values of normalized H$\alpha$ luminosity than most of the
slower rotators. At spectral type M9, the fastest rotators also show
weaker H$\alpha$ luminosity, but the saturation threshold may even be
as low as $v\,\sin{i} \approx 18$\,km\,s$^{-1}$. In the following
analysis of H$\alpha$ activity and rotation, we choose to exclude the
most rapid rotators ($v\sin{i} > 30$\,km\,s$^{-1}$).  We will come
back to the supersaturation effect in
Sect.\,\ref{sect:supersaturation}.  Unfortunately, this threshold
velocity is faster than can be accommodated by our measurement
technique for average magnetic fields. We therefore cannot measure
$Bf$ for the super-saturated stars.

The linear correlation coefficients between H$\alpha$-luminosity and
$v\,\sin{i}$ are 0.63, 0.32, and $-$0.34 for the M7, M8, and M9
subsamples, respectively (see Table\,\ref{tab:correlation}). If we
exclude M9 objects faster than $v\,\sin{i} = 18$\,km\,s$^{-1}$, the
coefficient is $-$0.13. The results of the Spearman rank tests clearly
show that clear evidence exists for a correlation between $\log{L_{\rm
    H\alpha}/L_{\rm bol}}$ and $v\,\sin{i}$ in the M7 subsample. On
the other hand, only weak indication exists for such a relation at M8,
but this hypothesis cannot be rejected either.  At M9, a correlation
between rotation and H$\alpha$ luminosity is very improbable in stars
slower than $v\,\sin{i} = 18$\,km\,s$^{-1}$, and if we include the
faster rotators, rotation and activity may even be anti-correlated,
but evidence for the latter is also weak. It is not surprising that
including the low-activity rapid rotators at spectral type M9 leads to
a weak anti-correlation. At this point, we cannot decide whether the
drop in H$\alpha$ luminosity around $v\,\sin{i} = 18$\,km\,s$^{-1}$ is
due to a super-saturation occuring at this threshold velocity, or
whether activity gradually weakens with faster rotation.

\begin{figure}
  \includegraphics[width=.45\hsize]{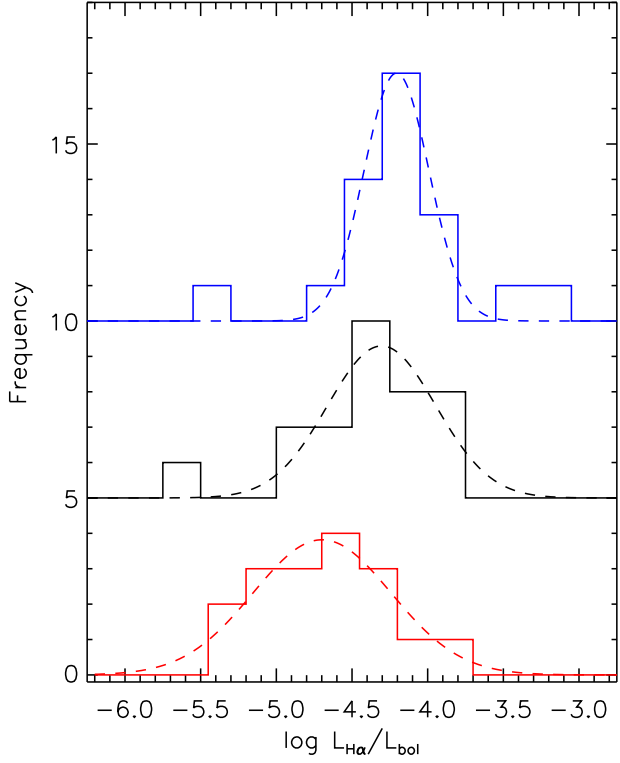}
  \caption{\label{fig:HistLHalpha} Histogram of normalized H$\alpha$
    luminosities for the three subsamples M7, M8, and M9 (top, center,
    and bottom row, respectively). Only objects with $v\,\sin{i} <
    30$\,km\,s$^{-1}$ are included. Gaussian fits to the distributions
    are overplotted as dashed lines, the mean values and standard
    deviations of the distributions are given in
    Table\,\ref{tab:Halpha}.}
\end{figure}

A well-known behaviour of the normalized H$\alpha$ luminosity is that
it becomes smaller with later spectral types.
Fig.\,\ref{fig:HistLHalpha} shows the distribution of $\log{L_{\rm
    H\alpha}/L_{\rm bol}}$ in the three subsamples of objects that are
rotating slower than $v\,\sin{i} = 30$\,km\,s$^{-1}$. If we assume
that the difference between H$\alpha$ luminosities within a spectral
bin are not due to different rotation velociteis but are purely
statistical, we might expect that $L_{\rm H\alpha}/L_{\rm bol}$, or
$\log{L_{\rm H\alpha}/L_{\rm bol}}$, simply scatters around a mean
value that depends on spectral type. In this case, the distribution of
the normalized H$\alpha$ luminosity could be described by a Gaussian
fit to the data.  We have tried a Gaussian fit to the distributions of
$\log{L_{\rm H\alpha}/L_{\rm bol}}$ in the three subsamples; the
results are overplotted in Fig.\,\ref{fig:HistLHalpha}. Note that the
choice of $\log{L_{\rm H\alpha}/L_{\rm bol}}$ (instead of $L_{\rm
  H\alpha}/L_{\rm bol}$) has no physical motivation. In all cases, the
values can consistently be described by a Gaussian distribution, which
shows that the scatter within a spectral bin may be purely random. 
Only two outliers have very low normalized activity levels exceeding
the Gaussian distributions (one each in the M7 and M8 subsamples).
Both are very slow rotators that probably lie on the unsaturated part
of a rotation-activity relation similar to early-M dwarfs, i.e., a
relation with a saturation velocity of a few km\,s$^{-1}$. Two objects
of the M7 subsample show extraordinary strong H$\alpha$ emission, one
of them, 2MASS~J07522939+161215, was probably observed during a flare.
The object with the strongest H$\alpha$ emission in our sample,
2MASS\,0041353$-$562112, is a young M8.0 brown dwarf with detected Li.
The width of the H$\alpha$ line at 10\,\% of the total intensity is
200\,km\,s$^{-1}$, i.e., much of the observed H$\alpha$ emission is
probably due to accretion \citep{White03}. This object is studied
separately in \citet{RE09}.

\begin{deluxetable}{lccc}
  \tablecaption{\label{tab:Halpha}Mean and width of the normalized
    H$\alpha$ luminosity distribution in stars with $v\,\sin{i} <
    30$\,km\,s$^{-1}$. }
  \tablewidth{0pt} 
  \tablehead{ & \colhead{M7} & \colhead{M8} & \colhead{M9\tablenotemark{a}}} 
  \startdata 
  Mean & $-4.16$ & $-4.30$ & $-$4.75 ($-$4.71) \\
  1 $\sigma$ width & \phantom{$-$}0.21 & \phantom{$-$}0.36 & \phantom{$-$}0.46 (\phantom{$-$}0.42) \\
  \enddata
  \tablenotetext{a}{Numbers in parentheses are excluding M9 objects
    faster than $v\,\sin{i} = 18$\,km\,s$^{-1}$ (see text).}
\end{deluxetable}

We conclude that the scatter in H$\alpha$ activity is mostly due to an
intrinsic, purely statistical scatter that is independent of rotation
and can be described by a Gaussian distribution. Outliers can probably
be explained by very slow rotation, flaring, or accretion. We
summarize the parameters of the Gaussian fits to the $\log{L_{\rm
    H\alpha}/L_{\rm bol}}$ distribution in the three subsamples in
Table\,\ref{tab:Halpha}, and we overplot the 1\,$\sigma$ range as grey
shaded areas in the top panel of Fig.\,\ref{fig:ActRot}. While the
mean level of H$\alpha$ activity is decreasing towards later spectral
type, the width of the distribution is growing, covering a larger area
in the $\log{L_{\rm H\alpha}/L_{\rm bol}}$ -- $v\,\sin{i}$ diagram.
Thus, objects like the two M8.0 stars at $v\,\sin{i} \approx
10$\,km\,s$^{-1}$ and $\log{L_{\rm H\alpha}/L_{\rm bol}} \approx -5$
may occupy the low end of the scatter among saturated objects instead
of an unsaturated branch extended to higher rotation. In other words,
we see no evidence for a higher saturation velocity at very low masses
but larger scatter and lower mean values of saturated activity among
ultracool dwarfs.

The two main conclusions of this part are that the rotation-activity
relation does not remain positive above $v\,\sin{i} =
$20--30\,km\,s$^{-1}$ in any spectral type, and that there is not a
tight correlation between velocity and activity in the velocity range
we can resolve.

\subsubsection{Rotation and Magnetic Field}

Average magnetic fields, $Bf$, of stars with $v\,\sin{i} <
20$\,km\,s$^{-1}$ are plotted as a function of $v\,\sin{i}$ in the
lower panel of Fig.\,\ref{fig:ActRot}. The linear correlation
coefficients between $Bf$ and $v\,\sin{i}$ are 0.32, 0.69, and -0.09
for the M7, M8, and M9 subsamples, respectively. The Spearman rank
test shows that there is weak (1$\sigma$) evidence for a correlation
between the average magnetic field and rotation among the M7 and M8
subsamples, and there is probably no correlation at M9. The strongest
fields observed in our sample, average magnetic fields on the order of
4\,kG, are observed only around $v\,\sin{i}$ of 10\,km\,s$^{-1}$. At
least among the M7 and M8 subsamples, very low magnetic fields below
our detection threshold are only seen in slowly rotating stars (the
one M9 object with no detectable field is a young brown dwarf, see
above). If we take away the highest and the lowest magnetic field
measurement in each subsample, the correlation coefficients are 0.24,
0.12, and 0.11 ($r_s = 0.25$, 0.12, and 0.21 with probabilities 0.37,
0.73, and 0.46) for the samples M7, M8, and M9, respectively.  This
means that the correlations between $Bf$ and $v\,\sin{i}$, that may
exist in the M7 and M8 subsamples, are mainly driven by two extreme
values in each subsample. Other than that, the scatter in $Bf$ is very
large and no clear correlation is found between magnetic field
strength and rotation.  This indicates that at M7 and M8, very low
magnetic fields still can only exist at very slow rotation, and that
the strongest fields require some rotation, but that magnetic field
generation is not as strongly tied to rotation as in earlier-type M
dwarfs.

\citet{Reiners09} showed the relation between average magnetic field
strength and rotation for a (non volume-limited) sample of 24 M-stars
with spectral types between M2 and M6. In their Fig.\,5, a clear
relation between $v\,\sin{i}$ and $Bf$ is visible: All stars rotating
at a detectable rate, i.e., faster than $\approx 3$\,km\,s$^{-1}$,
show average magnetic fields stronger than $\approx 1500$\,G, and nine
stars rotating slower than the detection threshold have fields below
1000\,G. The conclusion for the early- to mid-M stars is that a
rotation-magnetic field relation exists in the sense that all stars
rotating at a detectable rate produce strong magnetic fields, and that
no rapid rotators with weak fields are found. In the lower panel of
Fig.\,\ref{fig:ActRot}, we have highlighted the region of detectable
rotation and relatively weak magnetic fields (hatched area); this
region is not populated in stars of spectral type $<$M7 \citep[note
that][show that this correlation probably holds for earlier stars as
well]{Reiners09}. At spectral type M7, there are already quite a few
stars populating this area. At M8, field strengths in stars rotating
between $v\,\sin{i} = 5$ and 10\,km\,s$^{-1}$ are perhaps even weaker.
Obviously, there is a marked change in the magnetic field strengths
produced in rotating stars. This change occurs at spectral type M6/M7.

In ultracool dwarfs, rotational braking is weaker than in earlier type
stars \citep{Reiners08}, and our sample contains only very few objects
rotating at a very low rate. There is some evidence that very slow
rotators, at least in the M7 and M8 subsamples, can have very weak
fields. At detectable rotation, however, field generation is not as
efficient as in earlier-type stars. While there may be a weak
correlation between rotation and the weakest magnetic fields generated
at spectral types M7 and M8, the relation probably breaks down
altogether at spectral type M9. The main conclusion is that the
saturation of magnetic field generation, observed in sun-like stars
down to mid-M spectral types, breaks down around spectral type M7, and
the efficiency of magnetic field generation probably depends on
parameters other than rotation.

\subsubsection{H$\alpha$-Activity and Magnetic Field}

As we have seen in the last two Sections, the correlation between
average magnetic flux at M7 and M8 and rotation on one side, and
between H$\alpha$ luminosity and rotation on the other side, are
rather weak. We now ask the question whether chromospheric emission
observed in H$\alpha$ is still caused by magnetic flux, in other
words, whether H$\alpha$ luminosity still correlates with magnetic
fields in ultracool dwarfs. The linear correlation coefficients
between H$\alpha$ luminosity and $Bf$ for M7, M8, and M9 are 0.77,
0.66, and 0.61, respectively. The Spearman rank test excludes the null
hypothesis of no correlation at the 2$\sigma$ level at all spectral
types. This means that for the entire M dwarf range, there is evidence
for a correlation between H$\alpha$ luminosity and average magnetic
field.

\begin{figure}
  \plotone{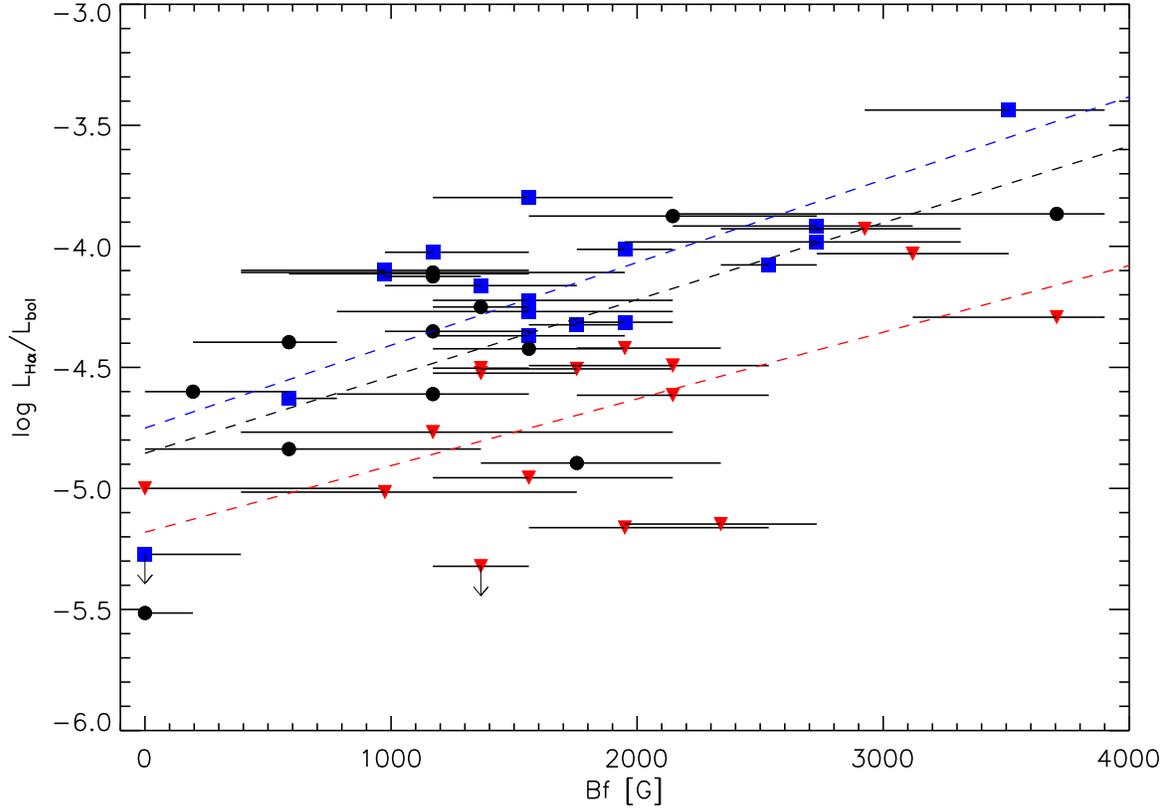}
  \caption{\label{fig:HalphaBf}Normalized H$\alpha$ luminosity as a
    function of average magnetic field, $Bf$. The three subsamples M7,
    M8, and M9 are shown with different symbols and colors (blue
    squares -- M7; black circles -- M8; red triangles -- M9). A linear
    fit was applied to each subsample, they are overplotted as dashed
    lines.}
\end{figure}

We show in Fig.\,\ref{fig:HalphaBf} the relation between H$\alpha$
luminosity and average magnetic field. The three subsamples M7, M8,
and M9 are shown with different colors and symbols, linear fits for
the three subsamples are overplotted. In all three subsamples, there
is a linear trend between H$\alpha$ and magnetic fields, higher
magnetic fields produce more H$\alpha$ emission. The correlations
between magnetic field and H$\alpha$ luminosity have similar strength
in all three subsamples, but they are offset with respect to each
other so that cooler objects in general produce less H$\alpha$
emission at the same magnetic field strength. This is probably
explained by the growing neutrality of the atmosphere which makes the
coupling between the magnetic fields and the atmosphere less efficient
at lower temperature \citep{Meyer99, Mohanty02}.

Fig.\,\ref{fig:HalphaBf} can be compared to Fig.\,8 in
\citet{Reiners09}, where a similar plot is shown for spectral types
M2--M6. At earlier spectral types, there is evidence for a linear
relation between magnetic field strength and $\log{L_{H\alpha}/L_{\rm
    bol}}$ for field strengths in the range 0--2\,kG. Above 2\,kG,
H$\alpha$ emission seems to be saturated and does not grow with higher
fields. From Fig.\,\ref{fig:HalphaBf}, it is difficult to assess
whether a similar saturation may occur in ultracool dwarfs because the
scatter even among the individual subsamples is substantial, and only
very few objects occupy the region of fields higher than 3\,kG. We
conclude that H$\alpha$ emission still depends on magnetic field
strength in a way that is similar to earlier stars. The scatter among
H$\alpha$ emission at constant field strength is substantial, it is
larger than at earlier spectral types, which may be due to differences
in atmospheric temperature and ionization. It is currently unclear
whether a saturation of H$\alpha$ emission occurs around field
strengths of 2\,kG like in earlier stars; more observations of stars
with very strong fields ($>3$\,kG) are required to answer that
question.

\subsection{Evidence for supersaturation?}
\label{sect:supersaturation}

Late-M dwarfs have small radii, relatively short rotation periods, and
probably large convective overturn times. While this renders a study
of activity difficult at Rossby numbers on the order of 1
\citep[weakly active or inactive stars, see][]{Pizzolato03}, it opens
the opportunity to investigate the other extreme of the
rotation-activity connection, i.e., to probe activity at very low
Rossby numbers where so-called ``supersaturation'' may exist. This
effect may be a result of coronal stripping due to centrifugal forces
\citep{Jardine04} or it could be due to some effect within the dynamo
itself.

To calculate the Rossby number, $Ro = P/\tau_{\rm conv}$, we require
knowledge of the rotation period $P$ and the convective overturn time
$\tau_{\rm c}$.  Unfortunately, no detailed calculations of the
convective overturn time for all M spectral types are available, and
it is even unclear where in the convection zone an effective overturn
time should be defined. Furthermore, M dwarf rotation periods are
difficult to measure. A sizeable number of periods is reported in
\cite{Kiraga07}, but still, for most M dwarfs, no rotation periods are
available. On the other hand, we have a large sample of $v\,\sin{i}$
measurements, from which we now estimate the rotation period to
investigate chromospheric activity at very low Rossby numbers.

In order to convert the projected rotational velocity $v\,\sin{i}$ to
rotation period for our late-M sample, we first estimate the radii of
the stars. Note that this only provides $Ro/\sin{i}$, which is an
upper limit of $Ro$.  To determine the radius, we employ the
mass-luminosity relation from \citet{Delfosse00} and the mass-radius
relation at an age of 5\,Gyrs from \cite{Baraffe98}. To compute the
masses, we use J-magnitudes from \cite{2MASS} and distances given in
Paper~I. We augment our sample by including the M0--M9 stars with
$v\,\sin{i}$ measurements from \citet{Delfosse98}, \citet{Mohanty03},
and \citet{RB07}. For them, we estimate the radius from a spectral
type-radius relation based on the values reported in \citet{Allen},
and we use $R = 0.1$\,R$_{\odot}$ for spectral type M8 and later.

We calculate the convective overturn time as a function of mass. We
estimate this function from a fit to the empirical results presented
in \citet{Kiraga07}, but we limit the convective overturn time to a
maximum of $\tau_{\rm conv} = 70$\,d consistent with the values given
for M dwarfs in \citet{Saar01}, which are taken from
\citet{Gilliland86}. The relation is $\tau_{\rm conv} [{\rm d}] = 86.9
- 94.3 M/{\rm M}_{\odot}$ for spectral type $<$M7 and $\tau_{\rm conv}
= 70$\,d for M7--M9. For the stars not contained in our new M7--M9
sample, the values for the mass are simply estimated to be the same as
the radii in solar units \citep[e.g., 0.2\,R$_\odot$ corresponds to
0.2\,M$_\odot$; see][]{Demory09}. Although there are certainly more
accurate ways to estimate radius and mass, this rather simple approach
is sufficient for our more qualitative investigation of this sample,
and the main uncertainty still resides in the value of the convective
overturn time. For example, if we were using the convective overturn
times from \citet{Kiraga07} for our late-M sample (i.e., without
limiting $\tau_{\rm c}$ to a maximum of 70\,d), the Rossby numbers
would be about 0.15\,dex smaller for the coolest stars, which is
equivalent to an uncertainty in the Rossby number of 30\,\%. This
difference does not affect our conclusions, and errors in the radius
estimate are likely smaller than this.

\begin{figure}
  \plotone{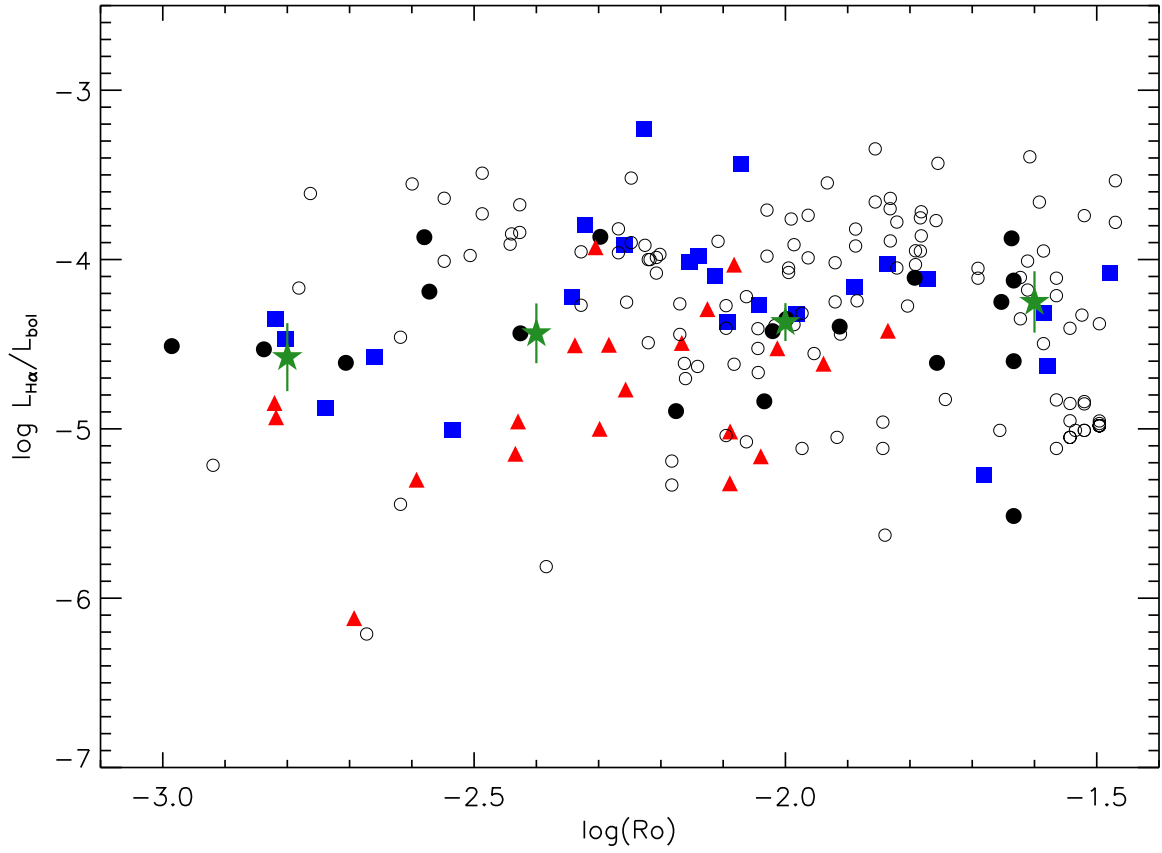}
  \caption{\label{fig:HalphaRossby}Normalized H$\alpha$ luminosity
    plotted as a function of Rossby number for our sample (blue
    squares -- M7; black circles -- M8; red triangles -- M9) and
    results taken from \citet{Delfosse98}, \citet{Mohanty03}, and
    \citet{RB07} (open circles). Green stars mark the median values
    when the sample is binned into four chunks, the error bars show
    the standard error of the median within each bin.}
\end{figure}

We plot the normalized H$\alpha$ luminosity as a function of Rossby
number in Fig.\,\ref{fig:HalphaRossby}; only M dwarfs with detected
H$\alpha$ emission are plotted. The M dwarfs occupy the range between
$-3.0$ and $-1.4$ in Rossby number, which is well within the
saturation regime \citep[see, e.g.,][]{Pizzolato03}. There is no
obvious difference between our subsamples of spectral types M7--M9 and
the other objects. However, there is a hint of a lack of very active
objects ($\log{L_{\rm H\alpha}/L_{\rm bol}} > -4.3$) with very low
Rossby numbers ($Ro < -2.6$). None of the objects in our late-M sample
occupies this region while many objects with higher Rossby numbers are
more active. Nine objects with $Ro < -2.6$ show relatively little
activity ($\log{L_{\rm H\alpha}/L_{\rm bol}} < -4.3$). This reflects
the evidence for supersaturation that we saw in
Fig.\,\ref{fig:ActRot}. From the earlier M dwarfs, however, one star,
G165-08 \citep[M4.5, $v\,\sin{i} =
55.5$\,km\,s$^{-1}$;][]{Delfosse98}, has a Rossby number smaller than
$-2.6$ and does show stronger activity. We divided the combined sample
of all M dwarfs into four bins of equal size in $Ro$ (0.4\,dex) and
calculated the median and its standard error for each bin.  These
values are overplotted in Fig.\,\ref{fig:HalphaRossby}, they show a
slight trend of lower H$\alpha$ luminosity with lower Rossby number,
but the trend remains inconclusive because of the large scatter and
the low number of objects particularly in the bin with the lowest
$Ro$. The Spearman rank correlation yields a coefficient of 0.20, and
the null hypothesis of no correlation can be rejected only at the
1$\sigma$-level.

M dwarfs are very variable. The highest activity levels are probably
connected with very strong flaring, i.e., the most active objects may
be observed during a temporal maximum of their activity. Although we
see no clear evidence for a supersaturation effect in our sample, the
lack of very active objects with very small $Ro$ -- at least among the
late-M dwarfs -- gives a hint towards the effect of supersaturation.
A plausible explanation for such an effect may be that coronal
stripping inhibits the occurence of strong flares. More observations
of very rapidly rotating M dwarfs ($v\,\sin{i} \ga 50$\,km\,s$^{-1}$)
would be helpful to settle this issue. In order to decide whether it
is the magnetic field that is weaker at very low Rossby number, or
whether it is the reduced radius of the corona, the best next step
would be to find a way to actually measure the magnetic fields on the
stars with very low Rossby numbers.

\section{Summary and Conclusions}

We investigated rotation, chromospheric emission, and average magnetic
fields in a volume-limited sample of 63 ultracool M dwarfs. The main
results from the analysis are the following: (1) The mean level of
H$\alpha$ luminosity is decreasing with lower temperature, and the
scatter among H$\alpha$ luminosity is larger at lower temperature, but
this scatter is not related to rotation; (2) we see a hint for
supersaturation occuring somewhere around a Rossby number of $-2.5$,
this may be caused by inhibited flaring due to centrifugal forces but
it could also be due to dynamo changes that we can't access until a
method for measuring fields in very rapid rotators is found, and in
any case requires confirmation from more observations; (3) in
ultracool dwarfs of a given temperature, H$\alpha$ luminosity still is
a function of magnetic field strength, it is unclear whether the
depedence shows saturation at high magnetic fields (as in earlier
stars); (4) a few very slowly rotating ultracool dwarfs have very weak
magnetic fields, but ultracool dwarfs that rotate at a detectable rate
generate substantial fields (with one exception that is a young brown
dwarf); (5) the saturation of magnetic field generation breaks down at
spectral type M7, before M7, rapid rotation always implies a field
strength on the order of 2\,kG or higher, while at M7 and later, stars
rotating as rapidly as 10\,km\,s$^{-1}$ can have magnetic fields on
the order of 1\,kG.

The lack of saturation of magnetic fields in ultracool dwarfs at
rotation velocities above $\approx 3$\,km\,s$^{-1}$ probably means
that the dynamo efficiency suffers a change around spectral type M7.
Before M7, rotation rates faster than $P \approx$ 2--3\,d causes
magnetic fields of 2\,kG strength or stronger. At M7 and later, the
fields can be much weaker. A possible explanation for this behavior is
that there occurs a change in the dynamo mechanism that predominantly
generates the magnetic fields we observe. The reason for such a change
is not at all clear. These stars do not harbor a tachocline, the
boundary layer between the radiative core and the convective shell,
which is believed to play an important role in the solar dynamo
\citep[e.g.,][]{Ossendrijver03}. However, the tachocline is believed
to disappear around spectral type M3 \citep[e.g.,][]{Siess00}, and
M4--M6 stars should already be completely convective. So far, we see
no obvious change in structure that could occur around spectral type
M6/M7.

Our understanding of purely turbulent dynamos is only at its
beginning, but large progress was made during the last years
\citep[e.g.,][]{Durney93, Dobler06, Browning08}. The observations of
ultracool dwarfs shown in this paper present empirical information on
fully convective dynamos, but future investigations of dynamo-relevant
parameters are required to put more constraints on dynamo modes that
are realized in stars.


\acknowledgements

We thank the referee for a very constructive and helpful report. Based
on observations collected at the European Southern Observatory,
Paranal, Chile, PIDs 080.D-0140 and 081.D-0190, and observed from the
W.M. Keck Observatory, which is operated as a scientific partnership
among the California Institute of Technology, the University of
California and the National Aeronautics and Space Administration. We
would like to acknowledge the great cultural significance of Mauna Kea
for native Hawaiians and express our gratitude for permission to
observe from atop this mountain.  A.R. has received research funding
from the DFG as an Emmy Noether fellow (RE 1664/4-1).  G.B. thanks the
NSF for grant support through AST06-06748.

\clearpage

\begin{deluxetable}{lcrrrrr}
  \tablecaption{\label{tab:results}Results}
  \tablewidth{0pt}
  \tablehead{\colhead{2MASS designation} & \colhead{SpType} & \colhead{$v\,\sin{i}$} & $\log{Ro}$ & H$\alpha$ EqW & \colhead{log($\frac{L_\mathrm{H\alpha}}{L_\mathrm{bol}}$)} & \colhead{$Bf$}\\
   & & \colhead{[km/s]} & & \colhead{[\AA]} & & \colhead{[G]}}
 \startdata
 $ 0435161-160657$ & M7.0 &     --\phantom{$\pm 2.0$} &       -- &      6.4 & $  -4.28$ &        --\phantom{$1^{800}_{800}$}  \\[1pt]
$ 0440232-053008$ & M7.0 &             $16.5\pm 2.0$ & $ -2.32$ &     19.3 & $  -3.80$ &              $1600^{+ 600}_{- 400}$ \\[1pt]
$ 0741068+173845$ & M7.0 &             $10.0\pm 2.0$ & $ -2.11$ &      9.7 & $  -4.10$ &              $1000^{+ 600}_{- 600}$ \\[1pt]
$ 0752239+161215$ & M7.0 &             $ 9.0\pm 2.0$ & $ -2.07$ &     44.4 & $  -3.44$ &              $3500^{+ 400}_{- 600}$ \\[1pt]
$ 0818580+233352$ & M7.0 &             $ 4.5\pm 2.0$ & $ -1.77$ &      9.4 & $  -4.11$ &              $1000^{+ 600}_{- 400}$ \\[1pt]
$ 1048126-112009$ & M7.0 &          $\le 3.0\pm 2.0$ & $ -1.58$ &      2.9 & $  -4.63$ &              $ 600^{+ 200}_{- 200}$ \\[1pt]
$ 1356414+434258$ & M7.0 &             $14.0\pm 2.0$ & $ -2.26$ &     14.8 & $  -3.92$ &              $2700^{+ 400}_{- 600}$ \\[1pt]
$ 1456383-280947$ & M7.0 &             $ 5.0\pm 2.0$ & $ -1.84$ &     11.5 & $  -4.02$ &              $1200^{+ 400}_{- 200}$ \\[1pt]
$ 1534570-141848$ & M7.0 &             $10.0\pm 2.0$ & $ -2.15$ &     11.8 & $  -4.01$ &              $2000^{+ 200}_{- 200}$ \\[1pt]
$ 0041353-562112$ & M7.5 &             $22.0\pm 2.0$ & $ -2.23$ &     92.7 & $  -3.23$ &        --\phantom{$1^{800}_{800}$}  \\[1pt]
$ 0148386-302439$ & M7.5 &             $48.0\pm 5.0$ & $ -2.82$ &      7.0 & $  -4.35$ &        --\phantom{$1^{800}_{800}$}  \\[1pt]
$ 0331302-304238$ & M7.5 &          $\le 3.0\pm 2.0$ & $ -1.58$ &      7.6 & $  -4.31$ &              $2000^{+ 200}_{- 200}$ \\[1pt]
$ 0351000-005244$ & M7.5 &             $ 6.5\pm 2.0$ & $ -1.89$ &     10.8 & $  -4.16$ &              $1400^{+ 400}_{- 400}$ \\[1pt]
$ 0417374-080000$ & M7.5 &             $ 7.0\pm 2.0$ & $ -1.98$ &      7.4 & $  -4.32$ &              $1800^{+ 200}_{- 200}$ \\[1pt]
$0429184-312356A$ & M7.5 &          $\le 3.0\pm 2.0$ & $ -1.48$ &     13.1 & $  -4.08$ &              $2500^{+ 200}_{- 200}$ \\[1pt]
$ 1006319-165326$ & M7.5 &             $16.0\pm 2.0$ & $ -2.34$ &      9.4 & $  -4.22$ &              $1600^{+ 600}_{- 400}$ \\[1pt]
$ 1155429-222458$ & M7.5 &             $33.0\pm 3.0$ & $ -2.66$ &      4.1 & $  -4.58$ &        --\phantom{$1^{800}_{800}$}  \\[1pt]
$ 1246517+314811$ & M7.5 &             $ 3.5\pm 2.0$ & $ -1.68$ & $<$  0.8 & $< -5.27$ &   $< 400$\phantom{$^{+800}_{-800}$} \\[1pt]
$ 1253124+403403$ & M7.5 &             $ 8.0\pm 2.0$ & $ -2.04$ &      8.4 & $  -4.27$ &              $1600^{+ 600}_{- 800}$ \\[1pt]
$ 1332244-044112$ & M7.5 &             $ 9.0\pm 2.0$ & $ -2.09$ &      6.7 & $  -4.37$ &              $1600^{+ 400}_{- 400}$ \\[1pt]
$ 1507277-200043$ & M7.5 &             $64.0\pm 6.0$ & $ -2.80$ &      5.3 & $  -4.47$ &        --\phantom{$1^{800}_{800}$}  \\[1pt]
$ 1521010+505323$ & M7.5 &             $40.0\pm 4.0$ & $ -2.74$ &      2.1 & $  -4.88$ &        --\phantom{$1^{800}_{800}$}  \\[1pt]
$ 1546054+374946$ & M7.5 &             $10.0\pm 2.0$ & $ -2.14$ &     16.3 & $  -3.98$ &              $2700^{+ 600}_{- 800}$ \\[1pt]
$ 1757154+704201$ & M7.5 &             $33.0\pm 3.0$ & $ -2.54$ &      1.5 & $  -5.01$ &        --\phantom{$1^{800}_{800}$}  \\[1pt]
$ 0019262+461407$ & M8.0 &             $68.0\pm10.0$ & $ -2.99$ &      5.9 & $  -4.51$ &        --\phantom{$1^{800}_{800}$}  \\[1pt]
$0027559+221932A$ & M8.0 &             $56.0\pm 6.0$ & $ -2.84$ &      5.7 & $  -4.53$ &        --\phantom{$1^{800}_{800}$}  \\[1pt]
$ 0123112-692138$ & M8.0 &             $26.0\pm 3.0$ & $ -2.57$ &     12.4 & $  -4.19$ &        --\phantom{$1^{800}_{800}$}  \\[1pt]
$ 0248410-165121$ & M8.0 &          $\le 3.0\pm 2.0$ & $ -1.65$ &     10.8 & $  -4.25$ &              $1400^{+ 200}_{- 200}$ \\[1pt]
$ 0320596+185423$ & M8.0 &             $15.0\pm 4.5$ & $ -2.30$ &     26.1 & $  -3.87$ &              $3700^{+ 200}_{-1600}$ \\[1pt]
$ 0517376-334902$ & M8.0 &             $ 8.0\pm 2.0$ & $ -2.02$ &      7.2 & $  -4.42$ &              $1600^{+ 400}_{- 400}$ \\[1pt]
$ 0544115-243301$ & M8.0 &          $\le 3.0\pm 2.0$ & $ -1.63$ &     14.4 & $  -4.12$ &              $1200^{+ 200}_{- 200}$ \\[1pt]
$ 1016347+275149$ & M8.0 &          $\le 3.0\pm 3.0$ & $ -1.64$ &     25.6 & $  -3.87$ &              $2100^{+ 600}_{- 600}$ \\[1pt]
$ 1024099+181553$ & M8.0 &             $ 7.5\pm 2.0$ & $ -2.03$ &      2.8 & $  -4.84$ &   $<1400$\phantom{$^{+800}_{-800}$} \\[1pt]
$1121492-131308A$ & M8.0 &             $27.0\pm 3.0$ & $ -2.58$ &     26.0 & $  -3.87$ &        --\phantom{$1^{800}_{800}$}  \\[1pt]
$ 1141440-223215$ & M8.0 &             $10.0\pm 2.0$ & $ -2.18$ &      2.4 & $  -4.90$ &              $1800^{+ 600}_{- 400}$ \\[1pt]
$ 1309218-233035$ & M8.0 &             $ 7.0\pm 2.0$ & $ -2.00$ &      8.5 & $  -4.35$ &              $1200^{+ 400}_{- 200}$ \\[1pt]
$ 1440229+133923$ & M8.0 &          $\le 3.0\pm 2.0$ & $ -1.63$ &      4.8 & $  -4.60$ &   $< 600$\phantom{$^{+800}_{-800}$} \\[1pt]
$ 1843221+404021$ & M8.0 &             $ 5.0\pm 3.2$ & $ -1.79$ &     15.0 & $  -4.11$ &              $1200^{+ 800}_{- 800}$ \\[1pt]
$ 2037071-113756$ & M8.0 &          $\le 3.0\pm 2.0$ & $ -1.63$ &      0.6 & $  -5.51$ &   $< 200$\phantom{$^{+800}_{-800}$} \\[1pt]
$ 2206227-204706$ & M8.0 &             $24.0\pm 2.0$ & $ -2.43$ &      7.0 & $  -4.44$ &        --\phantom{$1^{800}_{800}$}  \\[1pt]
$ 2306292-050227$ & M8.0 &             $ 6.0\pm 2.0$ & $ -1.91$ &      7.7 & $  -4.40$ &              $ 600^{+ 200}_{- 400}$ \\[1pt]
$ 2349489+122438$ & M8.0 &             $ 4.0\pm 2.0$ & $ -1.76$ &      4.7 & $  -4.61$ &              $1200^{+ 400}_{- 400}$ \\[1pt]
$ 2351504-253736$ & M8.0 &             $36.0\pm 4.0$ & $ -2.71$ &      4.7 & $  -4.61$ &        --\phantom{$1^{800}_{800}$}  \\[1pt]
$0024442-270825B$ & M8.5 &             $ 9.0\pm 2.0$ & $ -1.94$ &      5.5 & $  -4.62$ &              $2100^{+ 400}_{- 400}$ \\[1pt]
$ 0306115-364753$ & M8.5 &             $18.0\pm 2.0$ & $ -2.43$ &      2.5 & $  -4.96$ &              $1600^{+ 600}_{- 400}$ \\[1pt]
$ 1124048+380805$ & M8.5 &             $ 7.5\pm 2.0$ & $ -2.04$ &      1.6 & $  -5.16$ &              $2000^{+ 600}_{- 400}$ \\[1pt]
$ 1403223+300754$ & M8.5 &             $10.0\pm 2.0$ & $ -2.17$ &      7.3 & $  -4.49$ &              $2100^{+ 400}_{- 600}$ \\[1pt]
$ 1835379+325954$ & M8.5 &             $44.0\pm 4.0$ & $ -2.82$ &      3.2 & $  -4.85$ &        --\phantom{$1^{800}_{800}$}  \\[1pt]
$ 2226443-750342$ & M8.5 &             $15.0\pm 2.0$ & $ -2.34$ &      7.1 & $  -4.51$ &              $1800^{+ 400}_{- 400}$ \\[1pt]
$ 2331217-274949$ & M8.5 &             $ 9.0\pm 2.0$ & $ -2.08$ &     21.3 & $  -4.03$ &              $3100^{+ 400}_{- 400}$ \\[1pt]
$ 2353594-083331$ & M8.5 &             $ 4.5\pm 2.0$ & $ -1.84$ &      8.7 & $  -4.42$ &              $2000^{+ 400}_{- 200}$ \\[1pt]
$ 0019457+521317$ & M9.0 &             $ 9.0\pm 2.0$ & $ -2.13$ &     13.7 & $  -4.29$ &              $3700^{+ 200}_{- 600}$ \\[1pt]
$ 0109511-034326$ & M9.0 &             $13.0\pm 2.0$ & $ -2.28$ &      8.4 & $  -4.50$ &              $1400^{+ 200}_{- 200}$ \\[1pt]
$ 0334114-495334$ & M9.0 &             $ 8.0\pm 2.0$ & $ -2.09$ & $<$  1.3 & $< -5.32$ &              $1400^{+ 200}_{- 200}$ \\[1pt]
$ 0339352-352544$ & M9.0 &             $26.0\pm 3.0$ & $ -2.59$ &      1.3 & $  -5.30$ &        --\phantom{$1^{800}_{800}$}  \\[1pt]
$ 0443376+000205$ & M9.0 &             $13.5\pm 2.0$ & $ -2.30$ &      2.7 & $  -5.00$ &   $<1000$\phantom{$^{+800}_{-800}$} \\[1pt]
$ 0853362-032932$ & M9.0 &             $13.5\pm 2.0$ & $ -2.31$ &     31.8 & $  -3.93$ &              $2900^{+ 400}_{- 600}$ \\[1pt]
$ 1048147-395606$ & M9.0 &             $18.0\pm 2.0$ & $ -2.43$ &      1.9 & $  -5.15$ &              $2300^{+ 400}_{- 400}$ \\[1pt]
$ 1224522-123835$ & M9.0 &             $ 7.0\pm 2.0$ & $ -2.01$ &      8.0 & $  -4.52$ &              $1400^{+ 400}_{- 200}$ \\[1pt]
$ 1411213-211950$ & M9.0 &             $44.0\pm 4.0$ & $ -2.82$ &      3.1 & $  -4.93$ &        --\phantom{$1^{800}_{800}$}  \\[1pt]
$ 0024246-015819$ & M9.5 &             $33.0\pm 3.0$ & $ -2.69$ & $<$  0.2 & $< -6.12$ &        --\phantom{$1^{800}_{800}$}  \\[1pt]
$ 1438082+640836$ & M9.5 &             $12.0\pm 2.0$ & $ -2.26$ &      5.4 & $  -4.77$ &              $1200^{+1000}_{- 800}$ \\[1pt]
$ 2237325+392239$ & M9.5 &             $ 8.0\pm 2.0$ & $ -2.09$ &      3.1 & $  -5.02$ &              $1000^{+ 800}_{- 600}$ \\[1pt]

 \enddata
 \tablecomments{2MASS~J$0435161-160657$ is a double-lined
   spectroscopic binary (SB2), no rotation and magnetic flux analysis
   could be carried out, and the H$\alpha$ equivalent width was
   measured in the total spectrum including both components.}
\end{deluxetable}

\clearpage

\clearpage


\begin{thebibliography}{}
\bibitem[Baraffe et al., 1998]{Baraffe98}Baraffe, I., Chabrier, G.,
  Allard, F., \& Hauschildt, P.H., 1998, \aap, 337, 403
\bibitem[Browning, 2008]{Browning08}Browning, M.K., 2008, \apj, 676,
  1262
\bibitem[Burrows et al., 1997]{Burrows97}Burrows, A., Hubbard, W.B.,
  Saumon, D., \& Lunine, J.I., 1997, \apj, 406, 158
\bibitem[Cox, 2000]{Allen} Cox, A.N., 2000, Allen's astrophysical
  quantities, 4th ed., AIP Press, Springer, New York,~Editedy by
  Arthur N.~Cox
\bibitem[Crifo et al., 2005]{Crifo05}Crifo, F., Phan-Bao, N., Delfosse, X.,
  Forveille, T., Guibert, J., Mart\'in, E.L., \& Reyle\'e C., 2005,
  \aap, 441, 653
\bibitem[Cruz et al., 2003]{Cruz03}Cruz, K.L., Reid, I.N., Liebert, J.,
  Kirkpatrick, J.D., \& Lowrance, P.J., 2003, \aj, 126, 2421
\bibitem[Cruz et al., 2007]{Cruz07}Cruz, K.L., Reid, I.N., Kirkpatrick, J.D.,
  et al., 2007, \aj, 133, 439
\bibitem[Cutri et al., 2003]{2MASS} Cutri et al., 2003, The 2MASS
  All-Sky Catalog of Point Sources, University of Massachusetts and
  Infrared Processing and Analysis Center; IPAC/California Institute
  of Technology
\bibitem[Delfosse et al., 1998]{Delfosse98}Delfosse, X., Forveille,
  T., Perrier, C., \& Mayor, M., 1998, \aap, 331, 581
\bibitem[Delfosse et al., 2000]{Delfosse00}Delfosse, X., Forveille,
  T., S\'egransan, D., Beuzit, J.-L., Udry, S., Perrier, C., \& Mayor,
  M., 2000, \aap, 364, 217
\bibitem[Delfosse et al., 2001]{Delfosse01}Delfosse, X., Forveille,
  T., Mart\'in, E.L., et al., 2001, \aap, 366, L13
\bibitem[Demory et al., 2009]{Demory09}Demory, B.-O., S\'egransan, D.,
  Forveille, R., Queloz, D., Beuzit, J.-L., Delfosse, X., et al.,
  2009, \aap, accepted, \texttt{arXiv:0906.0602}
\bibitem[Dobler et al., 2006]{Dobler06}Dobler, W., Stix, M.,
  Brandenburg, A., 2006, \apj, 638, 336
\bibitem[Durney et al., 1993]{Durney93}Durney, B.R., de Young, D.S.,
  \& Roxburgh, I.W., 1993, Solar Physics, 145, 207
\bibitem[Gilliland, 1986]{Gilliland86}Gilliland, R.L., 1986, \apj,
  300, 339
\bibitem[Golimowski et al., 2004]{Golimowski04}Golimowski, D.A.,
  Leggett, S.K., Marley, M.S., et al., 2004, \aj, 127, 3516
\bibitem[Jardine, 2004]{Jardine04}Jardine, M., 2004, \aap, 414, L5
\bibitem[Johns-Krull \& Valenti, 2000]{JKV00}Johns-Krull, C.M., \&
  Valenti, J.A.,, 2000, ASP Conf.Ser., 198, p.371
\bibitem[Kiraga \& St\c{e}pie\'n, 2007]{Kiraga07}Kiraga, M. \&
  St\c{e}pie\'n, K., 2007, AcA, 57, 149
\bibitem[Meyer \& Meyer-Hofmeister, 1999]{Meyer99} Meyer, F., \&
  Meyer-Hofmeister, E., 1999, \aap, 341, L23
\bibitem[Mohanty et al., 2002]{Mohanty02}Mohanty, S., Basri, G., Shu, F.,
  Allard, F., \& Chabrier, G. 2002, \apj, 571, 469
\bibitem[Mohanty \& Basri, 2003]{Mohanty03} Mohanty, S., \& Basri, G.,
  2003, \apj, 583, 451
\bibitem[Ossendrijver, 2003]{Ossendrijver03}Ossendrijver, M., 2003,
  A\&AR, 11, 287
\bibitem[Phan-Bao et al., 2006]{PhanBao06}Phan-Bao, N., Bessel, M.S.,
  Mart\'in, E.L., et al., 2006, \mnras, 366, L40
\bibitem[Phan-Bao \& Bessel, 2006]{PhanBaoBessel}Phan-Bao, N., \&
  Bessel, M.S., 2006, \aap, 446, 515
\bibitem[Pizzolato et al., 2003]{Pizzolato03}Pizzolato, N., Maggio,
  A., Micela, G., Sciortino, S., \& Ventura, P., 2003, \aap, 397, 147
\bibitem[Press et al., 1992]{Press92}Press, W.H., Teukolsky, S.A.,
  Vetterling, W.T., Flannery, B.P., 1992, ``Numerical Recipes in C'',
  Camb. Univ. Press, Chapter 15
\bibitem[Reid et al., 2002]{Reid02}Reid, I.N., Kirkpatrick, J.D.,
  Liebert, J., Gizis, J.E., Dahn, C.C., \& Monet, D.G., 2002, \aj,
  124, 519
\bibitem[Reiners, 2007]{Reiners07}Reiners, A., 2007, A\&A, 467, 259
\bibitem[Reiners, 2009]{RE09}Reiners, A., 2009, ApJ, 702, L119
\bibitem[Reiners \& Basri, 2006]{Reiners06}Reiners, A., \& Basri, G.,
  2006, \apj, 644, 497
\bibitem[Reiners \& Basri, 2007]{RB07}Reiners, A., \& Basri, G.,
  2007, \apj, 656, 1121
\bibitem[Reiners \& Basri, 2008]{Reiners08}Reiners, A., \& Basri, G.,
  2008, \apj, 684, 1390
\bibitem[Reiners et al., 2009a]{Reiners09}Reiners, A., Basri, G., \&
  Browning, M., 2009a, \apj, 692, 538
\bibitem[Reiners et al., 2009b]{RBC09}Reiners, A., Basri, G., \&
  Christensen, U.R., 2009b, \apj, 697, 373
\bibitem[Reiners \& Basri, 2009]{I}Reiners, A., \& Basri, G., 2009,
  ApJ, accepted, \texttt{arXiv:0909.4647} (Paper~I)
\bibitem[Saar, 2001]{Saar01}Saar, S.H., 2001, ASP Conf. Ser., 223, 292
\bibitem[Siess et al., 2000]{Siess00}Siess, L., Dufour, E., \&
  Forestini, M., 2000, A\&A, 358, 593
\bibitem[West et al., 2004]{West04}West, A.A., Hawley, S.L.,
Walkowicz, L.M., Covey, K.R., Silvestri, N.M., and 6 authors
2004, \aj, 128, 426
\bibitem[West et al., 2008]{West08}West, A.A., Hawley, S.L.,
  Bochanski, J.J., et al., 2008, \aj, 135, 785
\bibitem[White \& Basri, 2003]{White03}White, R.J., \& Basri, G.,
  2003, \apj, 582, 1109
\bibitem[Wielen, 1977]{Wielen77}Wielen, R., 1977, \aap, 60, 263

\end{thebibliography}
\end{document}